\documentclass[journal]{IEEEtran}
\usepackage{soul}
\usepackage{graphicx}
\usepackage{mathtools}
\usepackage{amsmath}
\usepackage{amsfonts}
\usepackage{array}
\newcommand{\PreserveBackslash}[1]{\let\temp=\\#1\let\\=\temp}
\newcolumntype{C}[1]{>{\PreserveBackslash\centering}p{#1}}
\newcolumntype{R}[1]{>{\PreserveBackslash\raggedleft}p{#1}}
\newcolumntype{L}[1]{>{\PreserveBackslash\raggedright}p{#1}}
\usepackage[usenames]{color}
\usepackage{colortbl,booktabs}
\usepackage{tabularx}
\usepackage{multicol}
\usepackage{booktabs}
\usepackage[Symbol]{upgreek}
\usepackage{subfigure}
\usepackage{bm}
\usepackage{url}
\usepackage[usenames,dvipsnames,svgnames,table]{xcolor}
\usepackage{diagbox}
\usepackage{cite}
\usepackage{balance}
\usepackage{times}

\usepackage{stfloats}
\usepackage{balance}

\newtheorem{property}{Property}

\newtheorem{example}{Example}
\usepackage[nonumberlist,acronym,shortcuts]{glossaries}
\makeglossaries
\makeindex
\newacronym{ADSL}{ADSL}{Asymmetric Digital Subscriber Line}
\newacronym{ATP}{ATP}{Aggregate Transmit Power}
\newacronym{BER}{BER}{Bit Error Ratio}
\newacronym{BPSK}{BPSK}{Binary Phase Shift Keying}
\newacronym{CCMC}{CCMC}{Continuous-input Continuous-output Memoryless Channel}
\newacronym{CP}{CP}{Cyclic Prefix}
\newacronym{CPE}{CPE}{Customer Premises Equipment}
\newacronym{CWDD}{CWDD}{Column-Wise Diagonal Dominance}
\newacronym{DCMC}{DCMC}{Discrete-input Continuous-output Memoryless Channel}
\newacronym{DPU}{DPU}{Distribution Point Unit}
\newacronym{DSL}{DSL}{Digital Subscriber Line}
\newacronym{DSLAM}{DSLAM}{Digital Subscriber Line Access Multiplexer}
\newacronym{DSM}{DSM}{Dynamic Spectrum Management}
\newacronym{LD}{LD}{Line Driver}
\newacronym{LLR}{LLR}{Log-Likelihood Ratio}
\newacronym{ML}{ML}{Maximum Likelihood}
\newacronym{MIMO}{MIMO}{Multiple-Input Multiple-Output}
\newacronym{NEXT}{NEXT}{Near-End Crosstalk}
\newacronym{FEXT}{FEXT}{Far-End Crosstalk}
\newacronym{PDF}{PDF}{Probability Distribution Function}
\newacronym{PSD}{PSD}{Power Spectral Density}
\newacronym{PSK}{PSK}{Phase Shift Keying}
\newacronym{QAM}{QAM}{Quadrature-Amplitude Modulation}
\newacronym{QPSK}{QPSK}{Quadrature Phase Shift Keying}
\newacronym{SISO}{SISO}{Soft-In Soft-Out}
\newacronym{STBC}{STBC}{Space Time Block Code}
\newacronym{SM}{SM}{Spatial Modulation}
\newacronym{VDSL}{VDSL}{Very high speed Digital Subscriber Line}
\newacronym{V-BLAST}{V-BLAST}{Vertical-Bell Laboratories Layered Space-Time}
\newacronym{ZF}{ZF}{Zero-Forcing}
\begin{document}
{
%\tableofcontents
\title{\huge \color{black}{{Energy Efficient Transmission Based on Grouped Spatial Modulation for upstream DSL Systems}}}
\author{Jiankang~Zhang,~\IEEEmembership{Senior Member,~IEEE,}
Chao~Xu,~\IEEEmembership{Member,~IEEE,}
Tong~Bai,~\IEEEmembership{Member,~IEEE,}
Fasong~Wang,
Shida~Zhong,
}								

\maketitle
\begin{abstract}
The digital Subscriber Line (DSL) remains an important component of heterogeneous networking, especially in historic city-centers, where using optical fibre is less realistic. Recently, the power consumption has become an important performance  metric in telecommunication due to the associated environmental issues.  In the recent bonding model, customer sites have been equipped with two/four copper pairs, which may be exploited for designing grouped spatial modulation (SM) aiming for reducing the power consumption and mitigating the stubborn crosstalk in DSL communications. Explicitly, we view the two pair copper pairs equipped for each user as a group and propose an energy efficient transmission scheme based on grouped SM strategy  for the upstream DSL systems, which is capable of reducing the power consumption of the upstream transmitters by activating a single copper line of each user. More especially,  in order to compensate for the potential bit-rate reduction imposed by reducing the number of activated lines, the proposed scheme implicitly delivers ``\textit{virtual bits}" via activating/deactivating the lines in addition to the classic modulation scheme. This is particularly beneficial in the DSL context, because the cross-talk imposed by activating several lines may swamp the desired signal. Furthermore, a pair of near-optimal soft turbo detection schemes are proposed for exploiting the unique properties of the DSL channel in order to eliminate the error propagation problem of SM detection routinely encountered in wireless channels. Both the attainable energy-efficiency and the achievable Bit Error Ratio (BER)  are investigated. Our simulation results demonstrate that the proposed group-based SM is capable of outperforming the vectoring scheme both in terms of its energy efficiency for all the examined loop lengths and transmit powers. Moreover, the proposed group-based SM is capable of outperforming the vectoring scheme both in terms of  its BER performance in lower frequencies and longer serviced DSL loop lengths. 

%Monte Carlo simulations demonstrate that the proposed grouped SM outperforms vectoring scheme and STBC in terms of 

\end{abstract}
\IEEEpeerreviewmaketitle
\section{Introduction}\label{S1}

\glspl{DSL} remain the most widespread fixed broadband data transmission medium around the world with more than 300 million users \cite{bai2019impulsive,oksman2010itu} due to its benefits for both the operators and for the subscribers \cite{acatauassu2012measurement,zhang2018differential}. Where high-cost fibre installations are unaffordable, the widespread fixed copper cable infrastructure is capable of providing a reliable solution for numerous broadband access applications. Traditionally, \gls{DSL} technology aims for increasing the achievable data rate \cite{mazzenga2017analytical,mazzenga2016log,bai2019performance}, whilst minimizing the outage probability of the subscribers \cite{guenach2010power,bai2017discrete}. However, due to the power-thirsty nature of immersive multimedia applications, the power consumption has recently become of prime importance in telecommunications. In our specific context, the cable-distribution unit has a limited heat dissipation capability as well as powering capability \cite{oksman2016itu}. Furthermore, there is an increased interest in environmental issues \cite{guenach2010power}. Hence, the energy efficiency is at the top of the agenda of the various standardization bodies \cite{wolkerstorfer2008dynamic}, and it is becoming one of the principal design criteria for future networks and their equipment \cite{baliga2011energy,amin2015energy}.

In this spirit, low-power modes were introduced into the \gls{ADSL} standard, but unfortunately they exhibited instability in the operator's network \cite{wolkerstorfer2008dynamic}. Further attempts were made by dynamically managing the spectrum for  energy minimization in \gls{DSL} \cite{song2002dynamic}. Instead of solely maximizing the bit rate by \gls{DSM}, Wolkerstorfer, {\it{et al.}} \cite{wolkerstorfer2008dynamic} formulated a global optimization problem for minimizing the \gls{ATP} and struck a trade-off between the energy-efficiency and the bit-rate. A unified rate maximization and \gls{ATP} minimization framework was also designed by Guenach {\it{et al.}} \cite{guenach2009power} in order to strike a compelling trade-off between the bit rate and power consumption. Furthermore,  the total power consumption of \glspl{LD} was incorporated into the objective function by Guenach {\it{et al.}} in \cite{guenach2009trading,guenach2013energy}, since the line drivers account for $46\%$ of the total power consumption of \gls{ADSL} 2 and $36\%$ of that in the \gls{VDSL}2.

The power-conscious design of transceivers may also be capable of improving energy efficiency for example by mitigating the interference. In this spirit, Marrocco {\it{et al.}} \cite{marrocco2011energy} investigated the vectoring scheme's \cite{cendrillon2006near} capability of reducing power consumption and the impact of channel estimation errors imposed on the power consumption was also analyzed. The classical vectoring scheme has to activate all available twisted pairs for achieving multiplexing gain, which is not energy efficient. Maes {\it{et al.}} \cite{maes2014energy} proposed to combine the discontinuous operation with the vectoring scheme in order to reduce the power potentially. By contrast, \gls{SM} \cite{mesleh2008spatial,di2014spatial} has been shown to be an energy efficient wireless transmission scheme by activating a single antenna, which is capable of significantly reducing the total power consumption \cite{stavridis2012energy} by implicitly exploiting the spatial resources offered by the antenna indices. 

Similarly to wireless systems, fixed broadband networks also exploit the classic \gls{MIMO} paradigm, since multiple pairs of wires are bundled together in the bonding transmission\cite{coomans2015xg}. Particularly, in several countries, customers sites have been provided with two copper pairs in many place in Europe, where one copper line is for dedicated voice service and the other copper line is for fax or legacy dial-up data access. However, coupling may rise in these bonding transmission due to expanding bandwidth. The coupling effect is also known as crosstalk in \gls{DSL} terminology, which significantly limits the achievable data rate \cite{vatalaro2016sub,galli2016plc} and thus it will also impair the energy efficiency of transmission. Reducing the number of activated lines will intuitively mitigate the crosstalk, but it also reduces the aggregate information rate of the cable due to using a limited subset of the available pairs. Hence, our objective in this paper is to reduce the power consumption and mitigate the cross-talk without totally wasting the inactivated lines. This might seem unrealistic at first sight, but we will demonstrate that this ambitious may be achieved by grouping the wirelines for group-based \gls{SM} transmission. 

To expound a little further, in order to maintain a sufficiently high capacity whilst reducing the power consumption, we design an energy efficient scheme with the aid of our grouped \gls{SM} design philosophy. Explicitly, the proposed grouped \gls{SM} views the two pair twisted wires equipped for each user as a group, and each group only activates a single pair of wires, whilst exploiting the beneficial feature of \gls{SM}, namely that extra implicit information is delivered by the indices of the activated wires. More specifically, the novel contributions of this paper are as follows:
\begin{enumerate}
\item [1)] We conceive group-based \gls{SM} by activating a single twisted pair from {\color{black}two pairs twisted wires} of a group, which eliminates the crosstalk imposed by the adjacent pair. However, the \textit{virtual bits} are delivered via the specific twisted pair indices activated in the different groups.

\item [2)] Furthermore, a pair of soft turbo-detection schemes are proposed. Explicitly, the \gls{CWDD} property \cite{cendrillon2006near} of the \gls{DSL} channel below 102.4 MHz facilitates the low-complexity single-channel-based \gls{SM} detection for achieving near-optimum performance without encountering the classic error propagation problem of \gls{SM} detection routinely encountered in wireless channels.

\item [3)] Furthermore, both the \gls{CCMC} and the \gls{DCMC} capacity are quantified and the energy efficiency is investigated. The \gls{BER} performance is investigated at a number of representative cut-off frequencies, which span from 26.975 MHz to 101.975 MHz. The total achievable throughput and the overall BER performance attained upon increasing the bandwidth are also investigated for characterizing the overall performance.
\end{enumerate}

Throughout this paper, $\mathbb{C}$ denotes the complex number field, bold
 fonts are used to denote matrices and vectors. The transpose and Hermitian
 transpose operators are denoted by $(\cdot )^{\rm T}$ and $(\cdot )^{\rm H}$,
 respectively. The inverse operation is
 denoted by $(\cdot )^{-1}$, while $\mathbb{E}\{\cdot\}$ 
 stands for the expectation operations.  

 The rest of this paper is organized as follows. Section~\ref{S2} describes the
 system model of {color{black}the multi-user upstream \gls{DSL} systems.} Section~\ref{S3} presents the proposed group-based \gls{SM} transmission scheme as well as our soft turbo detection schemes. Section~\ref{S4} is devoted to our energy efficiency investigation. In Section~\ref{S5}, we present our simulation results for investigating the achievable energy efficiency and the achievable \gls{BER}. Our conclusions are offered in Section~\ref{S6}.

\section{System model}\label{S2}

The paradigm shift to \gls{MIMO} aided \gls{DSL} communications allows us to conceive sophisticated \gls{MIMO} signal processing techniques, such as precoding and/or postcoding applied for the transmitter and receiver, respectively.
In this section, we consider {\color{black}the upstream transmission between the user and the \gls{DPU}. Assuming that there is $N$ upstream users and each of them are quipped with $M$ pairs of twisted copper lines,} which indicates that we have $L = MN$, where $M = 1$ represents the simplest scenario of each group having a single pair of twisted lines. We assume that the \gls{NEXT} has been completely canceled  \cite{zidane2013vectored}, hence the interference at the receiver is imposed by the \gls{FEXT}. {\color{black}Upon assuming that the \gls{CP} is sufficiently long for eliminating the inter-carrier interference, we can process the multi-tone signals on a per tone basis \cite{ginis2002vectored}.} 
  More particularly, the vector $\boldsymbol{Y}_{k} \in \mathbb{C}^{L \times 1}$ of received signals on {\color{black}$k$-th} tone consists of $N$ vectors of $\boldsymbol{Y}_{k,n} \in \mathbb{C}^{M \times 1}$, which is given by
\begin{align}\label{EQ1:received_signal_1}
	\boldsymbol{Y}_{k} = \sqrt{P_{t/k}}\boldsymbol{H}_{k}\boldsymbol{X}_{k} + \boldsymbol{W}_{k} ,
\end{align}
where $P_{t/k} = P_{t}/K$ is the power allocated to a single tone of a single channel, $P_{t}$ is the total power transmitted  over $K$ subcarriers allocated  to a single channel, $\boldsymbol{H}_{k} \in \mathbb{C}^{L \times L}$ is the {\color{black}$k$-th} tone \gls{DSL} channel matrix subjected to \gls{FEXT}, $\boldsymbol{X}_{k} \in \mathbb{C}^{L \times 1}$ is the transmitted signal vector on tone $k$ and $\boldsymbol{W}_{k}\in \mathbb{C}^{L \times 1}$ is the noise vector. The elements $W_{k,l}, l = 1,2,\cdots, L$ of the additive noise contributions $\boldsymbol{W}_{k}$ are i.i.d. and obey the distribution of $\mathcal{CN}\left(0,\sigma_{w}^{2}\right)$. More specifically, $\boldsymbol{H}_{k}$ and $\boldsymbol{X}_{k}$ can be expanded as
\begin{align}
\boldsymbol{H}_{k} =& \left[\begin{array}{cccc}
\boldsymbol{H}_{k,1}^{1}&\boldsymbol{H}_{k,1}^{2}&\cdots&\boldsymbol{H}_{k,1}^{N}\\
\boldsymbol{H}_{k,2}^{1}&\boldsymbol{H}_{k,2}^{2}&\cdots&\boldsymbol{H}_{k,2}^{N} \\
\vdots&\vdots&\cdots&\vdots \\
\boldsymbol{H}_{k,N}^{1}&\boldsymbol{H}_{k,N}^{2}&\cdots&\boldsymbol{H}_{k,N}^{N} \\
\end{array}\right] , \\
\label{EQ3:X_k}
\boldsymbol{X}_{k} =& \left[\boldsymbol{X}_{k,1}^{\rm T},\boldsymbol{X}_{k,2}^{\rm T},\cdots,\boldsymbol{X}_{k,N}^{\rm T}\right]^{\rm T} ,
\end{align}
respectively. Furthermore, {\color{black}$\boldsymbol{H}_{k,g}^{n}$} and $\boldsymbol{X}_{k,n}$ are given by
\begin{align}\label{EQ4:sub_MIMO_ln}
	\boldsymbol{H}_{k,g}^{n} =& \left[\begin{array}{cccc}
H_{k,g,1}^{n,1}&H_{k,g,1}^{n,2}&\cdots&H_{k,g,1}^{n,M}\\
H_{k,g,2}^{n,1}&H_{k,g,2}^{n,2}&\cdots&H_{k,g,2}^{n,M} \\
\vdots&\vdots&\cdots&\vdots \\
H_{k,g,M}^{n,1}&H_{k,g,M}^{n,2}&\cdots&H_{k,g,M}^{n,M} \\
\end{array}\right] , \\
\label{EQ5:x-group-n}
\boldsymbol{X}_{k,n} =& \left[X_{k,n,1},X_{k,n,2},\cdots,X_{k,n,M}\right]^{\rm T} ,
\end{align}
where $H_{k,n,m}^{n,m}, n = 1,2,\cdots,N, m = 1,2,\cdots,M$ is the direct channel coefficient of the $m$-th channel of group $n$, while $H_{k,g,m'_{g}}^{n,m_{n}}, g \neq n$ represents the crosstalk coupling on the $m'$-th channel of group $g$ imposed by the $m$-th wire line channel of group $n$. Explicitly, the superscript $m_{n}$ and the subscript $m'_{g}$ are given by $m_{n} = (n - 1)M + m, m'_{g} = (g - 1)M + m', m, m' = 1,2,\cdots,M$, whilst $g$ and $n$ indicate the group indices. Moreover, $H_{k,n,m'_{n}}^{n,m_{n}}$ is the {\color{black}intra-group-crosstalk} imposed by the lines within a group, while $H_{k,g,m'_{g}}^{n,m_{n}}, g \neq n$ is inter-group-crosstalk imposed by other groups \cite{zafaruddin2015performance}.

The received signal can be detected based on the \gls{ZF} criterion \cite{cendrillon2006near} as follows
\begin{align}
	\hat{\boldsymbol{X}_{k}} =& \frac{1}{\sqrt{P_{t/k}}}\boldsymbol{H}_{k}^{-1}\boldsymbol{Y}_{k} , \nonumber\\
	=& \boldsymbol{X}_{k} + \frac{1}{\sqrt{P_{t/k}}}\boldsymbol{H}_{k}^{-1}\boldsymbol{W}_{k} .
\end{align}

We re-number the row and column indices of $\boldsymbol{H}_{k}$ as $\alpha$ and $\beta$, respectively, for simplifying the notation, where the mapping relationship between $H_{k,\alpha}^{\beta}$ and $H_{k,l,m}^{n,m'}$ is given by
\begin{align}
	g =& \lfloor \frac{\alpha}{M} \rfloor + 1, \\
	m =& (\alpha - 1) \text{ Mod } M  + 1, \\
	n =& \lfloor \frac{\beta}{M} \rfloor + 1, \\
	m' =& (\beta -1) \text{ Mod } M  + 1, 
\end{align}
where $\lfloor x \rfloor$ is a \textit{floor} function, which returns {\color{black}the} maximum integer less than $x$, and $(x \text{ Mod } M)$ returns the remainder after division of $x$ by $M$. Furthermore, $g$ and $n$ represent the group indices, while $m$ and $m'$ represent the channel indices with in each group.

\section{Proposed Group-based spatial modulation for DSL communications}\label{S3}

\begin{figure*}[htbp!]
\begin{center}
 \includegraphics[width=1.0\textwidth]{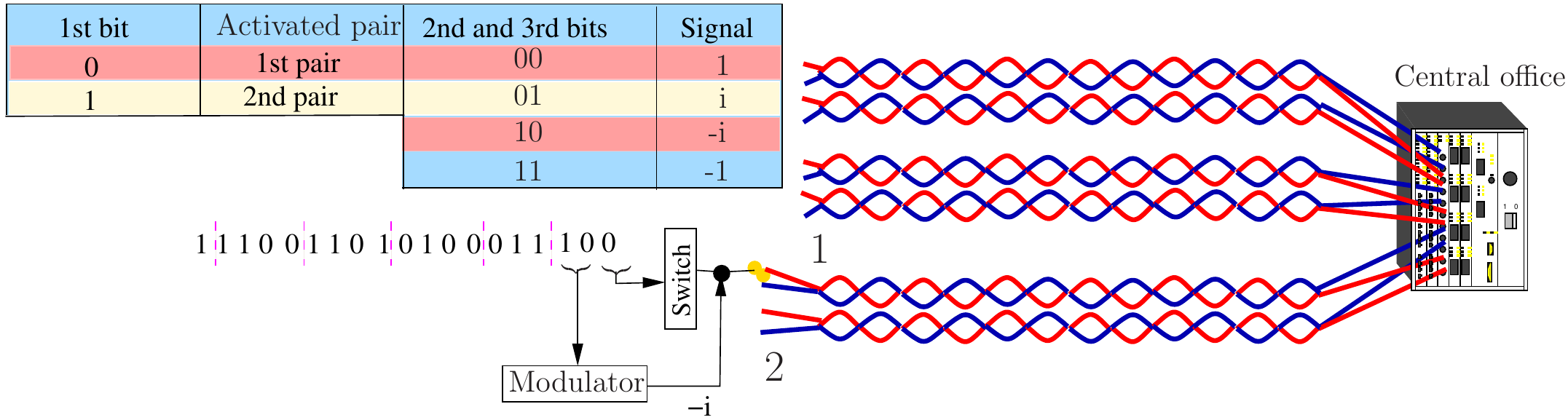}
\end{center}
\caption{An example of the proposed group-based SM for the upstream \gls{DSL} system, which supports $N = 3$ groups. Each group has $M = 2$ twisted pairs and the classical delivered signals are selected from the QPSK constellation.}
\label{FIG1:figure_for_SM_example}
\end{figure*}

Again, the grouped \gls{SM} scheme proposed supports $N$ upstream groups transmitted from the  \gls{DPU} to the central office. Each group has $M$ twisted pairs, but only a single pair is activated for transmission in each group, while the remaining $(M - 1)$ pairs remain inactive, hence we can implicitly transmit {\color{black}$p$} bits, where $p = \log_{2}\left(M\right)$. In this paper, we assume that $M$ is a  power of two. More specifically, the switch activates a pair using an ``{\it{activation pattern}}'' {\color{black}$\boldsymbol{u}_{k,n,m_{n}}$}, which is a $(M \times 1)$-element binary unit vector, while the rest of the elements in {\color{black}$\boldsymbol{u}_{k,n,m_{n}}$} are $0$. We define the set of activation patterns as $\mathbb{U}$, which consists of $M$ vectors. A logical `$1$' in an activation pattern indicates that the corresponding channel is active and `$0$' indicates that the corresponding channel is silent. Then, the activated channel transmits $q = \log_{2}\left(J\right)$ bits {\color{black}using} the classic modulation alphabet $\mathbb{A}$, where $J$ is the classic modulation order. For example, we have $q = 2$ for \gls{QPSK} and $q = 3$ for 8-\gls{QAM}. Therefore, the total number of bits conveyed by the proposed group-based \gls{SM} is given by
\begin{align}
	\eta = N\big[\log_{2}\left(M\right) + \log_{2}\left(J\right)\big] =& N[p + q],
\end{align}
where $N$ is the number of simultaneously supported upstream groups, and each group is equipped with $M$ twisted pairs.

The transmitted signal $\boldsymbol{X}_{k,n} \in \mathcal{C}^{M \times 1}$ of group $n$ in Eq.~(\ref{EQ5:x-group-n}) becomes
{\color{black}
\begin{align}\label{EQ14:X_k_n}
	\boldsymbol{X}_{k,n} =& \boldsymbol{u}_{k,n,m_{n}}\cdot s_{k,n}, m_{n} = 1,2,\cdots,M, n = 1,2,\cdots, N ,
\end{align}
}
where $s_{k,n} \in \mathbb{A}$ is a modulated signal selected from the classic \gls{QAM} constellation or \gls{PSK} constellation, while {\color{black}$\boldsymbol{u}_{k,n,m_{n}}$} represents the $m$-th channel activated for transmission. Furthermore, $\boldsymbol{X}_{k,n}$ is an element of $\mathcal{S}_{0}$ and there are $(JM)$ elements in $\mathcal{S}_{0}$.

Recalling Eq. (\ref{EQ3:X_k}) and Eq. (\ref{EQ14:X_k_n}), the transmitted signal vector {\color{black}$\boldsymbol{X}_{k} \in \mathcal{C}^{MN \times 1}$ in Eq.~(\ref{EQ1:received_signal_1}) consists of $\boldsymbol{X}_{k,n}, n = 1,2,\cdots, N$, which} can be rewritten as 
{\color{black}
\begin{align}
	\boldsymbol{X}_{k} =& \left[\boldsymbol{X}_{k,1}, \boldsymbol{X}_{k,2},\cdots,\boldsymbol{X}_{k,N},\right]^{\rm T}, m_{n} = 1,2,\cdots, M ,
\end{align}
}
where $\boldsymbol{X}_{k}$ is an element of $\mathcal{S}$ and there are $(JM)^{N}$ elements in $\mathcal{S}$.

%\subsection{Example: $N = 3$ groups, each of them deployed with a quad cable, saying $M = 4$, employing QPSK.}
\begin{example}
When $N = 3$ and $M = 2$, the number of bits that can be conveyed by the line activation pattern is $p = \log_{2}\left(M\right) = 1$ bits. The set of activation patterns is given by
\begin{align}
	\mathbb{U} =& \left\{\underbrace{\left[\begin{array}{c}0\\1\end{array}\right]}_{0},
	\underbrace{\left[\begin{array}{c}1\\0\end{array}\right]}_{1}
		\right\} ,
\end{align}
where the position of `1' in the vector represents the index of the activated channel.

The activated channels deliver \gls{QPSK} signals conveying the following bits
\begin{align}
	\mathbb{A} =& \left\{\underbrace{1}_{00}, \underbrace{i}_{01}, \underbrace{-1}_{11}, \underbrace{-i}_{10}\right\} .
\end{align}

As illustrated in Fig.~\ref{FIG1:figure_for_SM_example}, the first bit is input into the switcher of Fig.~\ref{FIG1:figure_for_SM_example}, which activates the corresponding channel according to the activation pattern. The following two bits are input into the modulator, which outputs \gls{QPSK} modulated signals.

\end{example}

\subsection{Hard detection for group-based SM over DSL}

At the receiver side, the optimal \gls{ML} principle can be employed for detecting the activated channels and the transmitted signals, which is formulated as
\begin{align}\label{EQ10:ML_detection}
	\left(\hat{\boldsymbol{m}}_{k},\hat{\boldsymbol{s}}_{k}\right) = \arg \min_{\boldsymbol{m} \in \mathbb{M}^{N},\boldsymbol{s} \in \mathbb{A}^{N}} \left\|\boldsymbol{Y_{k}} - \sqrt{P_{t/k}}\boldsymbol{H}_{k}\boldsymbol{X}_{k}\right\|_{2}^{2} ,
\end{align}
where $\mathbb{M} = \left\{1,2,\cdots,M\right\}$, $\boldsymbol{m} = \left[m_{1}, m_{2}, \cdots, m_{N}\right]^{\rm T}$ has $|\mathbb{M}|^{N} = M^{N}$ potential combinations for the activated channels, while $\boldsymbol{s} = \left[s_{1}, s_{2}, \cdots, s_{N}\right]^{\rm T}$ has $|\mathbb{A}|^{N} = J^{N}$ potential combinations of the transmitted signals delivered by $N$ groups. In order to acquire the optimal \gls{ML} detection results, we have to enumerate $M^{N} \times J^{N} = \left(JM\right)^{N}$ different combinations of the activated channels and of the classic transmitted signals, which is impractical for a large number of groups, while employing high-order modulation. 

Serendipitously, the \gls{DSL} channel subjected to crosstalk has the following \gls{CWDD} property \cite{cendrillon2006near} {\color{black}{ below 80 MHz, which has also demonstrated by our experimental observation.}}
\begin{property}[\textit{Column-wise diagonal dominance}]
The diagonal elements of the \gls{DSL} channel suffering from crosstalk have the largest magnitude on each column of $\boldsymbol{H}_{k}$, which may be mathematically formulated as
\begin{align}
	\left|H_{k,\alpha}^{\beta}\right| \le \left|H_{k,\alpha}^{\alpha}\right|, \forall \alpha \neq \beta .
\end{align}
\end{property}
 
Exploiting the \gls{CWDD} property of the \gls{DSL} channel, the computational complexity of the detection problem in Eq.~(\ref{EQ10:ML_detection}) can be significantly reduced by sequentially detecting the activated channels and the classic transmitted signals.

\begin{figure}[tbp!]
\begin{center}
 \includegraphics[width=1.0\columnwidth,angle=0]{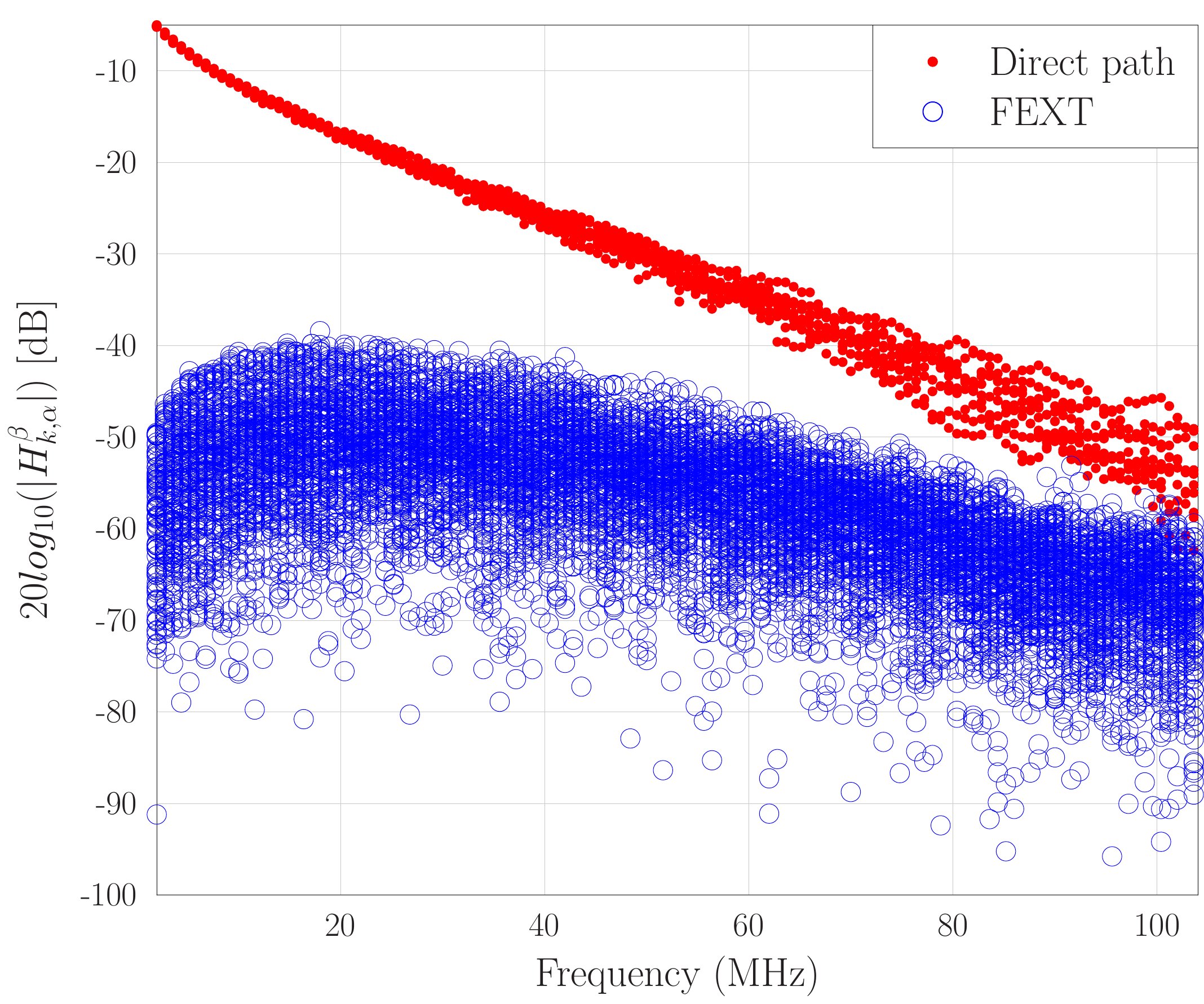}
\end{center}
\vspace{-6mm}
\caption{Measurement of the crosstalk channel of a 200 meter BT cable consisting of 10 twisted copper pairs.}
\label{FIG2:FEXT_channel}
\end{figure}

Actually, the \gls{CWDD} property of \gls{DSL} channel indicates that $\left|H_{k,\alpha}^{\beta}\right|^{2} \ll \left|H_{k,\alpha}^{\alpha}\right|^{2}, (n',m') \neq (n'',m'')$. This feature has also been verified by our the measured results in the \
Ultra-Fast Lab of BT Group plc., as shown in Fig.~\ref{FIG2:FEXT_channel}. Explicitly, we measured the frequency response
of a 200 meter BT cable consisting of 10 twisted copper pairs, where each wire has a diameter of 0.5 mm. The measurement results demonstrate the \gls{CWDD} property of \gls{DSL}, and the results also show a gradually decreasing ratio between the desired signal and the FEXT upon increasing the frequency.

{\color{black}
Upon assuming that the $m^{*}$-th line is activated by group $n^{*}$, we have {\color{black}$\alpha = (n^{*} - 1)M + m^{*}$}. Thus, 
\begin{align}
	\left|Y_{k,\alpha}\right|^{2} =& \left|\sum\limits_{\beta = 1}^{MN}\sqrt{P_{t,cl}}H_{k,\alpha}^{\beta}X_{k,\beta} + W_{k,\alpha}\right|^{2}, \nonumber \\
	\ge & \left|\sqrt{P_{t,cl}}H_{k,\alpha}^{\alpha}X_{k,\alpha}\right|^{2} \nonumber\\
	& - \left|\sqrt{P_{t,cl}}\sum\limits_{\beta = 1,\beta \neq \alpha}^{MN}H_{k,\alpha}^{\beta}X_{k,\beta} + W_{k,\alpha}\right|^{2}, \\
	= & \left|\sqrt{P_{t,cl}}H_{k,\alpha}^{\alpha}X_{k,\alpha}\right|^{2} - \nonumber\\ 
	& \left|\sum\limits_{n = 1,n \neq n^{*}}^{N}\sum\limits_{m = 1}^{M}\sqrt{P_{t,cl}}H_{k,\alpha}^{n,m_{n}}X_{k,n,m_{n}} + W_{k,\alpha}\right|^{2},\nonumber\\
	\ge & P_{t,cl}\left|H_{k,\alpha}^{\alpha}\right|^{2} - P_{t,cl}\sum\limits_{n = 1,n \neq n^{*}}^{N}\sum\limits_{m = 1}^{M}\left|H_{k,\alpha}^{n,m_{n}} \right|^{2} - \sigma_{w}^{2},
\end{align}

For $\alpha' = (n^{*} - 1)M + m, m \neq m^{*}, m = 1,2,\cdots,M$, $Y_{k,\alpha'}$ is given by
\begin{align}\label{EQ21}
	\left|Y_{k,\alpha'}\right|^{2} =& \left|\sum\limits_{\beta = 1}^{MN}\sqrt{P_{t,cl}}H_{k,\alpha'}^{\beta}X_{k,\beta} + W_{k,\alpha'}\right|^{2}, \nonumber\\
	= & \left|\sum\limits_{n = 1}^{N}\sum\limits_{m = 1}^{M}\sqrt{P_{t,cl}}H_{k,\alpha'}^{n,m_{n}}X_{k,n,m_{n}} + W_{k,\alpha'}\right|^{2},\nonumber\\
	\le & P_{t,cl}\sum\limits_{n = 1}^{N}\sum\limits_{m = 1}^{M}\left|H_{k,\alpha'}^{n,m_{n}} \right|^{2} + \sigma_{w}^{2},
	\end{align}
}

Considering all the other potential noise contributions, using an increased white noise \gls{PSD} of $-140$ dBm/Hz \cite{acatauassu2012measurement} is reasonable for \gls{DSL}, which indicates that the power of white noise is significantly lower than $P_{t/k}\left|H_{k,\alpha}^{\alpha}\right|^{2}$. Omitting the term of $\sigma_{w}^{2}$ in Eq.~(\ref{EQ21}) and exploiting the \gls{CWDD} property of the \gls{DSL} channel, we have 
\begin{align}
	&\left|Y_{k,\alpha}\right|^{2} - \left|Y_{k,\alpha'}\right|^{2} \!\!=\!\! P_{t/k}\left|H_{k,\alpha}^{\alpha}\right|^{2} \!\!-\!\! P_{t/k}\sum\limits_{n = 1,n \neq n^{*}}^{N}\sum\limits_{m = 1}^{M}\left|H_{k,\alpha}^{n,m_{n}} \right|^{2}\nonumber\\
	& -  P_{t/k}\sum\limits_{n = 1}^{N}\sum\limits_{m = 1}^{M}\left|H_{k,\alpha'}^{n,m_{n}} \right|^{2} > 0 .
\end{align}
Thus, $\left|Y_{k,\alpha}\right|^{2} > \left|Y_{k,\alpha'}\right|^{2}$.

Then, the activated channel of group $u$ can also be detected on the basis of the group as follows
\begin{align}\label{EQ23}
m_{n} = \arg \max_{m = \{1,2,\cdots,M\}} \left|Y_{k,n,m}\right|^{2} ,
\end{align}
where $m_{n}$ indicates that the $m_{n}$-th channel is activated by group $n$. It was demonstrated in \cite{xu2013spatial,jeganathan2008spatial} that Eq.~(\ref{EQ23}) may result in error propagation in wireless channels. By contrast, due to the \gls{CWDD} property of the \gls{DSL} channel, the single-line-based \gls{SM} detection of Eq.~(\ref{EQ23}) effectively reduces the $(MN)$ - variable problem-which is reminiscent of the wireless \gls{V-BLAST} detection problem- to an $N$-variable  detection problem without suffering from error propagation. This is rather appealing feature of employing \gls{SM} in \gls{DSL}.

Note that $\left|Y_{k,\alpha}\right|^{2}$, may be less than $\left|Y_{k,\alpha'}\right|^{2}$ when the DSL is very long or the frequency is very high, since the \gls{CWDD} property of the \gls{DSL} channel becomes less pronounced upon increasing the loop length or increasing the frequency, as seen in Fig.~\ref{FIG2:FEXT_channel}. This will result in a poor \gls{BER} performance due to mis-detection of the activated channels, which will be investigated in \textit{Section V}.

\begin{figure*}[htbp!]
\begin{center}
 \includegraphics[width=1.0\textwidth,angle=0]{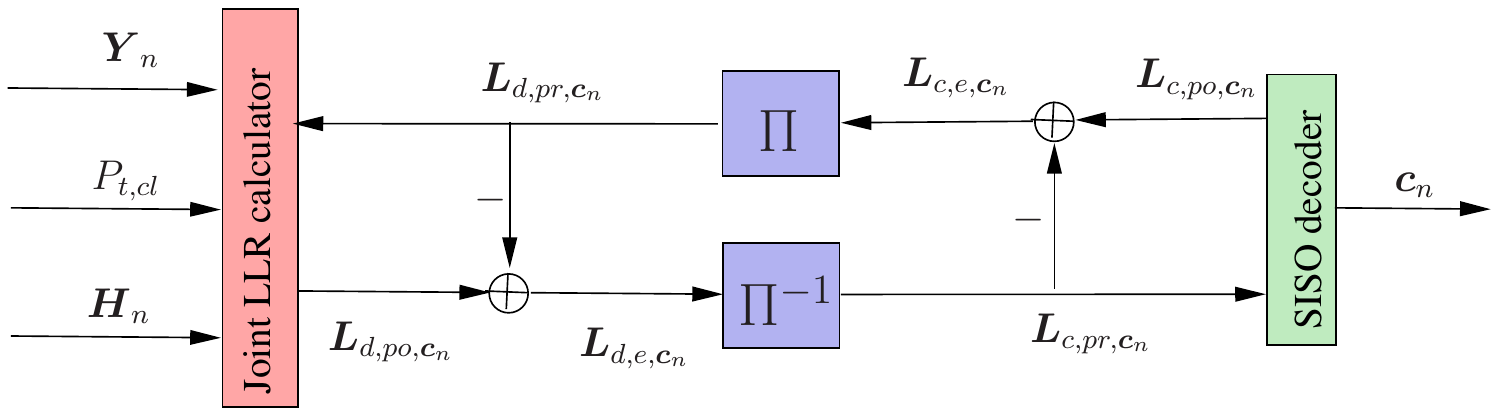}
\end{center}
\vspace{-6mm}
\caption{Schematic of the SOSD-I. The subscripts $d$ and $c$ of $\boldsymbol{L}$ are associated with the LLR calculator and
 channel decoder, respectively, while the subscripts $pr, po$ and $e$ are used for representing the {\it{a priori, a posteriori}} and extrinsic information. The tone index of $k$ is omitted in this figure for avoiding confusion.}
\label{FIG2:DS_SM_joint}
\end{figure*}

The activated channel indices are marked as a vector $\boldsymbol{m}_{k}^{*} = \left[m_{1},m_{2},\cdots,m_{N}\right]^{\rm T}$, where $m_{n} = (n - 1)M + m$. Then the received signal model of Eq.~(\ref{EQ1:received_signal_1}) can be truncated as 
\begin{align}
	\tilde{\boldsymbol{Y}}_{k} = \sqrt{P_{t/k}}\tilde{\boldsymbol{H}}_{k}\tilde{\boldsymbol{s}}_{k} + \tilde{\boldsymbol{W}}_{k} ,
\end{align}
where $\tilde{\boldsymbol{Y}}_{k} \in \mathcal{C}^{N \times 1}, \tilde{\boldsymbol{s}}_{k} \in \mathcal{C}^{N \times 1}$ and $\tilde{\boldsymbol{W}}_{k} \in \mathcal{C}^{N \times 1}$ only have the elements at the position $\boldsymbol{m}_{k}^{*}$  of the vectors $\boldsymbol{Y}_{k}, \boldsymbol{X}_{k}$ and $\boldsymbol{W}_{k}$, respectively. Furthermore, $\tilde{\boldsymbol{H}}_{k} \in \mathcal{C}^{N \times N}$ is a truncated version of $\boldsymbol{H}_{k}$, retaining its $\boldsymbol{m}_{k}^{*}$ rows and $\boldsymbol{m}_{k}^{*}$ columns.

Upon applying the \gls{ZF} criterion, we arrive at the linear crosstalk canceler \cite{cendrillon2006near} formulated as:
\begin{align}
	\widehat{\tilde{\boldsymbol{s}}}_{k} =& \frac{1}{\sqrt{P_{t/k}}}\tilde{\boldsymbol{H}}_{k}^{-1}\tilde{\boldsymbol{Y}}_{k} , \\
	=& \tilde{\boldsymbol{s}}_{k} + \frac{1}{\sqrt{P_{t/k}}}\tilde{\boldsymbol{H}}_{k}^{-1}\tilde{\boldsymbol{W}}_{k} , \\
	=& \tilde{\boldsymbol{s}}_{k} + \breve{\boldsymbol{W}}_{k} ,
\end{align}
where $\breve{\boldsymbol{W}}_{k} = \frac{1}{\sqrt{P_{t/k}}}\tilde{\boldsymbol{H}}_{k}^{-1}\tilde{\boldsymbol{W}}_{k}$ and $\breve{W}_{k,n}$ obeys the distribution of $\mathcal{CN}\left(0, \sigma_{\breve{w}_{k,n}}\right)$, while $\sigma_{\breve{w}_{k,n}}$ is given by
\begin{align}\label{EQ39:new_sigma}
	\sigma_{\breve{w}_{k,n}}^{2} =& \frac{1}{P_{t/k}}\left[\tilde{\boldsymbol{H}}_{k}^{-1}\left(\tilde{\boldsymbol{H}}_{k}^{-1}\right)^{\rm H}\right]_{n,n} \sigma_{w}^{2}.
\end{align}
In Eq.~(\ref{EQ39:new_sigma}), $[\cdot]_{n,n}$ represents the specific element on the $n$-th row and the $n$-th column. 

\subsection{Soft turbo detection}

In this subsection, we propose a pair of sub-optimal soft detection schemes for detecting both the active channels and the signals delivered by the active channels. The first scheme is termed as Sub-Optimal Soft Detection I (SOSD-I), which jointly detects the active channels and the classic symbols based on soft \textit{a posteriori} information, as seen in Fig.~\ref{FIG2:DS_SM_joint}. The second scheme is termed as Sub-Optimal Soft Detection II (SOSD-II), which separately detects the active channels by exploiting the \gls{CWDD} property of the \gls{DSL} channels and the classic symbols based on soft \textit{a posteriori} information, respectively, as seen in Fig.~\ref{FIG3:DS_SM_scheme2}. The soft information provided either by the SOSD-I or by the SOSD-II scheme is fed to a \gls{SISO} turbo decoder.

(1) \textit{Sub-Optimal Soft Detection I}

The $i$-th bit carried by the channel of group $n$ on tone $k$ is denoted as $a_{k,n}[i]$, which represents the activation pattern vector mapped by the $i$-th bit of group $n$ on tone $k$. For example, $[0\,0]$ is mapped to $[0\,0\,0\,1]^{\rm T}$, indicating that the first channel is activated, while the $i$-th bit carried by the classic symbol of the active channel of group $n$ on tone $k$ is denoted as $b_{k,n}[i]$, which represents the $M$-QAM data stream mapped by the $i$-th bit of group $n$ on tone $k$. For example $[0\,0\,0\,0]$ is mapped to $(-3 + 3j)$ in a system employing $16$-QAM. We introduce the mapping of $\Big[a_{k,n}[1]\,\cdots\,a_{k,n}[p]\,b_{k,n}[1]\,\cdots\,b_{k,n}[q]\Big]$ to  $\Big[c_{k,n}[1]\,c_{k,n}[2]\,\cdots\,c_{k,n}[(p + q)]\Big]$, where the first $p$ bits of $\boldsymbol{c}_{k,n}$ are carried by the indices of the active channels, while the remaining $q$ bits are carried by the signals delivered by the activated channels.

The \gls{LLR} calculator delivers  the \textit{a posteriori} information of the bit $c_{k,n}[i]$ expressed in terms of its \gls{LLR} as \cite{zhang2012turbo,zhang2014evolutionary}
{\color{black}	
\begin{align}\label{EQ:Po_LLR} % eq 24
L_{po,c_{k,n}[i]} =& \ln \frac{Pr\big[\boldsymbol{Y}_{k,n}\big|c_{k,n}[i]
 = 0\big]}{Pr\big[\boldsymbol{Y}_{k,n}\big|c_{k,n}[i] = 1\big]} + \ln
 \frac{Pr\left[c_{k,n}[i] = 0\right]}{Pr\left[c_{k,n}[i] = 1\right]} ,
  \nonumber \\ 
	=&  L_{e,c_{k,n}[i]} + L_{pr,c_{k,n}[i]} .	
\end{align}
}
The second term of Equation
  (\ref{EQ:Po_LLR}) may be denoted by $L_{pr,c_{k,n}[i]}$, which represents the {\it{a priori}}
  \glspl{LLR} of the interleaved and encoded bits $c_{k,n}[i]$. By contrast, the first term of Eq.~(\ref{EQ:Po_LLR}), which is denoted by $L_{e,c_{k,n}[i]}$ represents the extrinsic
  information delivered by the \gls{LLR} calculator of Fig.~\ref{FIG2:DS_SM_joint}. Explicitly, the extrinsic
  information $L_{e,c_{k,n}[i]}$ is calculated based on the received signal $\boldsymbol{Y}_{k,n}$ and on the {\it{a priori}} information concerning the encoded bits of group $n$, except for the $i$-th bit.
	
	More specifically, the soft \textit{a posteriori} information $L_{po,c_{k,n}[i]}$ associated with bit $b_{k,n}[i]$ can be approximated by the low-complexity Max-Log-MAP \cite{koch1990optimum} as
		\begin{align}
		\label{EQ:po_opt_ml_joint} % eq25
 L_{po,c_{k,n}[i]} =& \max_{\begin{subarray}{c}c_{k,n}[i]
 = 0 \\\forall \boldsymbol{X}_{k,n} \in \mathcal{S}_{0}\end{subarray}} d^{i} - \max_{\begin{subarray}{c}c_{k,n}[i]
 = 1 \\ \forall \boldsymbol{X}_{k,n} \in \mathcal{S}_{0}\end{subarray}} d^{i} ,
	\end{align}
	where the probability metric $d^{i}$ is given by
	\begin{align}\label{EQ31:d_i}
		d^{i}\!\!=\!\!& -\frac{\left\|\boldsymbol{Y}_{k,n}
 - \sqrt{P_{t/k}}\boldsymbol{H}_{k,n}^{n}\boldsymbol{X}_{k,n}\right\|_{2}^{2}}{2\sigma^{2}_{w}} \!\!+\!\! \sum\limits_{j = 1}^{(p + q)}\tilde{c}_{k,n}[j] L_{pr,c_{k,n}[j]} .
	\end{align}
	In Eq.~(\ref{EQ31:d_i}), $\Big[\tilde{c}_{k,n}[1]\,\cdots \,\tilde{c}_{k,n}[p + q]\Big] = \text{de2bi}(i)$ refers to the mapping of the \gls{SM} bits to the signal $\boldsymbol{X}_{k,n}$, where $\text{de2bi}(i)$ converts decimal integers to binary symbols.

(2) \textit{Sub-Optimal Soft Detection II}
\begin{figure*}[htbp!]
\begin{center}
 \includegraphics[width=1.0\textwidth,angle=0]{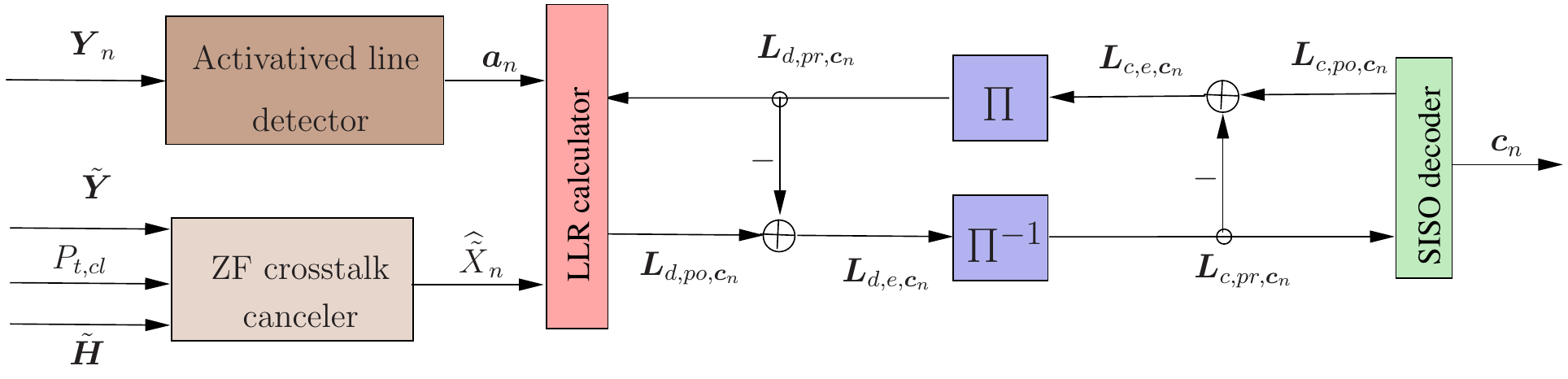}
\end{center}
\vspace{-6mm}
\caption{Schematic of the SOSD-II. The subscripts $d$ and $c$ of $\boldsymbol{L}$ are associated with the LLR calculator and
 channel decoder, respectively, while the subscripts $pr, po$ and $e$ are used for representing the {\it{a priori, a posteriori}} and extrinsic information. The tone index of $k$ is omitted in this figure for simplicity.}
\label{FIG3:DS_SM_scheme2}
\end{figure*}

By exploiting the \gls{CWDD} property of the \gls{DSL} channel, the activated channels may be separately detected according to  the hard detection rule of Eq.~(\ref{EQ23}). The bits carried by the indices of active channels are mapped to the \textit{a posteriori} information as 
\begin{align}
	\left\{\begin{array}{c}
	L_{po,a_{k,n}[i]} = -1 \quad \text{if} \quad a_{k,n}[i] = 0, \\
	L_{po,a_{k,n}[i]} = 1 \quad \text{if} \quad a_{k,n}[i] = 1 .
	\end{array}\right.
\end{align}

Furthermore, the \gls{LLR} calculator delivers the \textit{a posteriori} information of the bits $b_{k,n}[i]$ as	
\begin{align}
		\label{EQ:po_SOSD2} % eq25
 L_{po,b_{k,n}[i]} =& \max_{\begin{subarray}{c}b_{k,n}[i]
 = 0 \\\forall \boldsymbol{X}_{k,n} \in \mathcal{S}_{0}\end{subarray}} d^{i} - \max_{\begin{subarray}{c}b_{k,n}[i]
 = 1 \\ \forall \boldsymbol{X}_{k,n} \in \mathcal{S}_{0}\end{subarray}} d^{i} ,
\end{align}
where the probability metric $d^{i}$ of SOSD-II is given by
	\begin{align}\label{EQ45:d_i}
		d^{i} =& -\frac{\left|\widehat{\tilde{s}}_{k,n}
 - s_{k,n}\right|^{2}}{2\sigma^{2}_{w}} + \sum\limits_{j = 1}^{q}\tilde{b}_{k,n}[j] L_{pr,b_{k,n}[j]} .
	\end{align}
	In Eq.~(\ref{EQ45:d_i}), $\Big[\tilde{b}_{k,n}[1]\,\cdots \,\tilde{b}_{k,n}[(p + q)]\Big] = \text{de2bi}(i)$ represents the decimal to binary mapping of the signal $s_{k,n}$ delivered.

\section{Capacity and energy-efficiency analysis}\label{S4}
In this section, we investigate both the \gls{CCMC} and \gls{DCMC} capacities, where the \gls{CCMC} capacity is the maximum achievable rate of a given channel in case of Gaussion distributed transmitted signals. However, the \gls{DCMC} capacity may be  more practical for guiding the design of practical modulation schemes. Furthermore, the energy efficiency of the proposed group-based SM is investigated.

\subsection{CCMC Capacity analysis}
The \gls{CCMC} capacity of the \gls{MIMO} channel is achieved by maximizing the mutual information between the input signal and output signal per channel, which is given by 
\begin{align}
	C_{\text{CCMC}} = \max_{p(\boldsymbol{S})}\left[\frac{1}{L} \mathcal{H}(\boldsymbol{Y}) - \frac{1}{L} \mathcal{H}(\boldsymbol{Y}|\boldsymbol{S})\right] .
\end{align}

Thus, the \gls{CCMC} capacity on tone $k$ of a vectoring scheme \cite{cendrillon2006near} can be formulated as
{\color{black}
\begin{align}
	C_{k}^{\text{vec}} =& \mathbb{E}\left\{\Delta_{f}\log_{2}\det\left(\boldsymbol{I}_{L} + \frac{P_{t/k}^{\text{vec}}}{L\sigma_{w}^{2}}\boldsymbol{H}_{k}^{\rm H}\boldsymbol{H}_{k}\right)\right\} ,
\end{align}
}
where $L = MN$ is the total number of channels of $N$ users and $\Delta_{f}$ is the bandwidth of a single tone. $P_{t/k}^{\text{vec}}$ represents the power allocated to a single tone of a single channel for vectoring scheme.

Since \gls{SM} conveys its information both via the classic constellation points of the modulated signal and by the spatial domain in the form of the activated line index, its capacity is given by the sum of the classic signal domain capacity and  of the spatial domain capacity \cite{yang2008information}. Thus, the capacity of the proposed group-based \gls{SM} can be written as
\begin{align}\label{EQ36:CCMC_SM}
	C_{\text{SM,k,total}} =& \sum\limits_{n = 1}^{N}\left(C_{\text{signal},k,n} + C_{\text{spatial},k,n}\right) ,
\end{align}
where $C_{\text{signal},k,n}$ and $C_{\text{spatial},k,n}$ represent the capacity of the signal domain and the capacity of the spatial domain, respectively.

More explicitly, the first term of Eq.~(\ref{EQ36:CCMC_SM}) represents a single-input multiple-output (SIMO) system's capacity, which is maximized when the input is assumed to be a Gaussian-distributed continuous signal, formulated as:
\begin{align}
C_{\text{signal},k,n} =& \max I\left(s_{k,n};\boldsymbol{Y}_{k,n}|m\right) ,\nonumber\\
=& \frac{\Delta_{f}}{M}\sum\limits_{m = 1}^{M}\log_{2}\left(1 + \frac{P_{t/k}^{\text{SM}}\left\|\boldsymbol{H}_{k}^{n_{m}}\right\|_{2}^{2}}{\sigma_{w}^{2}}\right) ,
\end{align}
where $I\left(s_{k,n};\boldsymbol{Y}_{k,n}|m\right)$ is the mutual information between $s_{k,n}$ and $\boldsymbol{Y}_{k,n}$ conditioned on $m$, $P_{t/k}^{\text{SM}}$ represents the power allocated to a single tone of a single channel for group-based SM.
 
Similarly, the second term of Eq.~(\ref{EQ36:CCMC_SM}) is also maximized by the Gaussian \gls{PDF} of the output signal as:
\begin{align}
&p\left(\boldsymbol{Y}_{k,n}|m\right) = \frac{1}{\det \left(\pi \boldsymbol{R}_{\boldsymbol{Y}\boldsymbol{Y}|m}\right)}\exp\left(-\boldsymbol{Y}_{k,n}\boldsymbol{R}_{\boldsymbol{Y}\boldsymbol{Y}|m}\boldsymbol{Y}_{k,n}^{\rm H}\right) , \nonumber \\
&= \frac{1}{ \frac{\pi}{\sigma_{w}^{2}}\left(1 + \frac{P_{t/k}^{\text{SM}}\left\|\boldsymbol{H}_{k}^{n_{m}}\right\|^{2}}{\sigma_{w}^{2}}\right)}\times \nonumber\\
&  \exp\left(-\boldsymbol{Y}_{k,n}\left(P_{t/k}^{\text{SM}}\boldsymbol{H}_{k}^{n_{m}, \rm H}\boldsymbol{H}_{k}^{n_{m}} + \sigma_{w}^{2} \boldsymbol{I}_{M}\right)^{-1}\boldsymbol{Y}_{k,n}^{\rm H}\right) .
\end{align}

As a result, the second term of Eq.~(\ref{EQ36:CCMC_SM}) may be further expressed as
\begin{align}
	&C_{\text{spatial},k,n}= \max I\left(m;\boldsymbol{Y}_{k,n}\right) ,\nonumber \\
	\label{EQ39:C_spatial}
	&= \max_{p(m)}\int \int p(\boldsymbol{Y}_{k,n}|m)p(m)\log_{2}\left(\frac{p\left(\boldsymbol{Y}_{k,n}|m\right)}{P(\boldsymbol{Y}_{k,n})}\right) d m d \boldsymbol{Y}_{k,n} ,
	\end{align}
	where the average output PDF is given by $p\left(\boldsymbol{Y}_{k,n}\right) = \int p(\boldsymbol{Y}_{k,n}|m)p(m)d m$. Naturally, Eq.~(\ref{EQ39:C_spatial}) is maximized when the input PDF $p(m)$ is also Gaussian {\color{black} given that the power is constraint}. However, the index $m$ of the activated line is confined to the limited discrete range of $(1 \le m \le M)$, which cannot be generalized by allowing $M$ to tend to infinity. As a remedy, according to \cite{xin2006mimo}, the activated lines index $M$ may be interpreted as a discrete signal. Hence, Eq.~(\ref{EQ39:C_spatial}) is maximized for equi-probable source of $p(m) = \frac{1}{M}$ as 
	\begin{figure*}[tp!]\setcounter{equation}{39}
\begin{align}\label{EQ40:CCMC_SM_index}
	C_{\text{spatial},n}=& \frac{1}{M}\sum\limits_{m = 1}^{M} \mathbb{E}\left\{\log_{2}\frac{\frac{M}{1 + \frac{P_{t/k}^{\text{SM}}\left\|\boldsymbol{H}_{k}^{n_{m}}\right\|^{2}}{\sigma_{w}^{2}}}\exp\left(-\boldsymbol{Y}_{k,n}\left(P_{t/k}^{\text{SM}}\boldsymbol{H}_{k}^{n_{m}, \rm H}\boldsymbol{H}_{k}^{n_{m}} + \sigma_{w}^{2} \boldsymbol{I}_{M}\right)^{-1}\boldsymbol{Y}_{k,n}^{\rm H}\right)}{\sum\limits_{m' = 1}^{M}\frac{1}{1 + \frac{P_{t/k}^{\text{SM}}\left\|\boldsymbol{H}_{k}^{n_{m'}}\right\|^{2}}{\sigma_{w}^{2}}}\exp\left(-\boldsymbol{Y}_{k,n}\left(P_{t/k}^{\text{SM}}\boldsymbol{H}_{k}^{n_{m'}, \rm H}\boldsymbol{H}_{k}^{n_{m'}} + \sigma_{w}^{2} \boldsymbol{I}_{M}\right)^{-1}\boldsymbol{Y}_{k,n}^{\rm H}\right)}\right\} .
\end{align}	
\hrulefill
\vspace*{-4mm}
\end{figure*}

We note that Eq.~(\ref{EQ40:CCMC_SM_index}) was also invoked for estimating the capacity of Space-Shift Keying (SSK) and Space-Time Shift Keying (STSK) in \cite{jeganathan2009space} and \cite{sugiura2010coherent}, respectively. Owning to the discrete nature of the activation index, it was demonstrated in \cite{yang2015design} that SM suffers from a capacity loss compared to  the Bell Laboratories Layer Space-Time (BLAST) scheme. Nonetheless, we will demonstrate in \emph{Section \ref{S5a}} that in DSL systems SM is capable of outperforming the vectoring scheme\footnotemark\footnotetext{The vectoring scheme in DSL systems is equivalent to the BLAST scheme in wireless communication systems.} in terms of its energy efficiency quantified in terms of the  \gls{CCMC} capacity normalized by the power consumption in Mbps/J.

In summary, it can be seen in Eq.~(\ref{EQ36:CCMC_SM}) that the CCMC capacity of SM is higher  than that of its conventional SIMO counterpart, which was also observed in \cite{rajashekar2014reduced}.

\subsection{DCMC capacity analysis}
In practical digital telecommunication systems, the input signals are typically discrete \gls{QAM} or \gls{PSK} signals, instead of being continuous Gaussian-distributed. Moreover, the \gls{DCMC} capacity is directly determined by the choice of modulation, which can be adaptively controlled. This feature can be exploited for providing different-rate services to different users.

The \gls{DCMC} capacity of the \gls{MIMO} channel is given by \cite{xin2006mimo}
\begin{align}\label{capacity_DCMC_1}
	C \!\!=\!\!& \max_{p(\boldsymbol{S}_{i}) \in \mathcal{S}} \sum\limits_{i = 1}^{I}\int p(\boldsymbol{Y} | \boldsymbol{S}_{i}) p(\boldsymbol{S}_{i}) \log_{2}\frac{p(\boldsymbol{Y} | \boldsymbol{S}_{i})}{\sum\limits_{i = 1}^{I}p(\boldsymbol{Y} | \boldsymbol{S}_{i}) p(\boldsymbol{S}_{i})} d \boldsymbol{Y} ,
\end{align}
where $I$ is the number of signals in the set of $\mathcal{S}$, and $p(\boldsymbol{Y} | \boldsymbol{S}_{i})$ is given by
\begin{align}
	p(\boldsymbol{Y} | \boldsymbol{S}_{i}) =& \frac{1}{\left(\pi \sigma_{w}^{2}\right)^{N_{R}}}\exp\left(-\frac{\left\|\boldsymbol{Y} - \sqrt{P_{t/k}}\boldsymbol{H}\boldsymbol{S}_{i}\right\|_{2}^{2}}{\sigma_{w}^{2}}\right) .
\end{align}
Again, the maximum of Eq.~(\ref{capacity_DCMC_1}) is achieved, when the input signal vectors $\boldsymbol{S}_{i}$ are equiprobable, i.e., $p(\boldsymbol{S}_{i}) = \frac{1}{I}$ for $\forall i = 1,2,\cdots, I$. Then, the \gls{DCMC} capacity of the \gls{MIMO} channel quantified in Eq.~(\ref{capacity_DCMC_1}) can be rewritten as
\begin{align}
	C =& \frac{1}{I}\sum\limits_{i = 1}^{I} \mathbb{E} \left\{\log_{2}\frac{I p(\boldsymbol{Y} | \boldsymbol{S}_{i})}{\sum\limits_{i' = 1}^{I}p(\boldsymbol{Y} | \boldsymbol{S}_{i'})}\right\} .
\end{align}

Thus, the capacity on tone $k$ of the proposed grouped \gls{SM} can be written as
\begin{align}
	C_{k}^{\text{SM}} =& \frac{\Delta_{f}}{(JM)^{N}}\sum\limits_{i = 1}^{(JM)^{N}} \mathbb{E} \left\{\log_{2}\frac{(JM)^{N} p(\boldsymbol{Y}_{k}^{\text{SM}} | \boldsymbol{S}_{k,i}^{\text{SM}})}{\sum\limits_{i' = 1}^{(JM)^{N}}p(\boldsymbol{Y}_{k}^{\text{SM}} | \boldsymbol{S}_{k,i'}^{\text{SM}})}\right\} ,
\end{align}
where $K$ is the total number of tones, and again, $\Delta_{f}$ is the bandwidth of a single tone, while $p(\boldsymbol{Y}_{k} | \boldsymbol{S}_{k,i})$ is given by
\begin{align}\label{EQ45}
	p(\boldsymbol{Y}_{k}^{\text{SM}} | \boldsymbol{S}_{k,i}) \!=\!& \frac{1}{\left(\pi \sigma_{w}^{2}\right)^{MN}}\exp\left(\!\!-\frac{\left\|\boldsymbol{Y}_{k}^{\text{SM}} \!\!-\!\! \sqrt{P_{t/k}^{\text{SM}}}\boldsymbol{H}_{k}\boldsymbol{S}_{k,i}\right\|_{2}^{2}}{\sigma_{w}^{2}}\right) .
\end{align}

In Eq.~(\ref{EQ45}), $P_{t/k}^{\text{SM}}$ is the power allocated to a single tone of a single channel for the group-based SM scheme.

By contrast, the vectoring is a classic \gls{DCMC} \gls{MIMO} channel. Thus, its capacity on tone $k$ is given by
\begin{align}
	C_{k}^{\text{vec}} =& \frac{\Delta_{f}}{J^{N_{R}}}\sum\limits_{i = 1}^{J^{N_{R}}} \mathbb{E} \left\{\log_{2}\frac{J^{N_{R}} p(\boldsymbol{Y}_{k}^{\text{vec}} | \boldsymbol{S}_{k,i}^{\text{vec}})}{\sum\limits_{i' = 1}^{J^{N_{R}}}p(\boldsymbol{Y}_{k}^{\text{vec}} | \boldsymbol{S}_{k,i'}^{\text{vec}})}\right\} ,
\end{align}
where $N_{R} = MN$ represents the total number of received channels, $\boldsymbol{S}_{k,i} \in \mathcal{S}_{\text{vec}}$ and $\mathcal{S}_{\text{vec}}$ has $J^{N_{R}}$ elements.

\subsection{Power consumption model and energy efficiency analysis for DSL}
The total power consumption of a \gls{DSL} transceiver is constituted by that of the digital front-end, of the analog front-end and of the \gls{LD} \cite{wolkerstorfer2012modeling}. The \gls{LD} is responsible for outputting the transmit power $P_{t}$ required for signal transmission, thus we only consider the power consumption of the \gls{LD}, but neglect the power consumed by the digital front-end and the analog front-end in our ensuing energy efficiency analysis. There are various types \glspl{LD}, but here we consider the popular class-AB \gls{LD} power model, which can be characterized as \cite{wolkerstorfer2012modeling}
\begin{align}\label{EQ60:LD_power}
	P_{\text{LD}} =& V_{s}\left(I_{Q} + \sqrt{\frac{2}{\pi}\frac{P_{t}}{R_{\text{line}}^{'}}}\right) + P_{\text{Hybrid}} ,
\end{align}
where $V_{s}$ is the supply voltage of the \gls{LD}, $I_{Q}$ is the quiescent current, $R_{\text{line}}^{'}$ is the transformed resistance of the line, and $P_{\text{Hybrid}}$ is the power consumed by the hybrid circuit.

For the sake of a fair comparison, all schemes are allocated the same power $P_{t}^{\text{total}}$ for each group. Thus, the powers allocated to each of the active channels of the proposed group-based \gls{SM} and to the vectoring are $P_{t}^{\text{SM}} = P_{t}^{\text{total}}$, $P_{t}^{\text{vec}} = \frac{P_{t}^{\text{total}}}{M}$. Hence, the power consumption on each tone are $P_{t/k}^{\text{SM}} = \frac{P_{t}^{\text{SM}}}{K}$ and $P_{t/k}^{\text{vec}} = \frac{P_{t}^{\text{vec}}}{K}$ for the group-based SM and vectoring scheme, respectively.

Recalling that the proposed group-based SM only activates a single line in each group, while the vectoring activates all lines in each group, the energy efficiency of them {\color{black} on the $k$-th tone} can be formulated as
{\color{black}
\begin{align}
\label{EE_k_vec}
	\eta_{k}^{\text{vec}} = & \frac{C_{k}^{\text{vec}}}{N P_{\text{LD},k}^{\text{vec}}} , \\
	\label{EE_k_sm}
	\eta_{k}^{\text{SM}} = & \frac{C_{k}^{\text{SM}}}{N P_{\text{LD},k}^{\text{SM}}} .
\end{align}
where $P_{\text{LD},k}^{\text{vec}} = P_{\text{LD}}^{\text{vec}} / k$ and $P_{\text{LD},k}^{\text{SM}} = P_{\text{LD}}^{\text{SM}} / k$. And $P_{\text{LD}}^{\text{vec}}$ and $P_{\text{LD}}^{\text{SM}}$ are calculated using Eq.~{\ref{EQ60:LD_power}} with the supplied transmit power of $P_{t}^{\text{vec}}$ and $P_{t}^{\text{SM}}$. Furthermore, the total energy efficient of all tones are given by
\begin{align}
	\eta^{\text{vec}} = & \frac{\sum\limits_{k = 1}^{K}C_{k}^{\text{vec}}}{NP_{\text{LD}}^{\text{vec}}} , \\
	\eta^{\text{SM}} = & \frac{\sum\limits_{k = 1}^{K}C_{k}^{\text{SM}}}{NP_{\text{LD}}^{\text{SM}}} .
\end{align}
More especially, the energy efficient of the proposed group-based SM and the vectoring scheme are attained by replacing $C_{k}^{\text{SM}}$ and $C_{k}^{\text{vec}}$ with their corresponding CCMC capacity and DCMC capacity, respectively.
}

Note that both the \gls{SM}'s capacity is higher than that of a single \gls{DSL} channel, but  lower than that of the vectoring's capacity. However, \gls{SM} only activates a single pair of twisted lines, which significantly reduces the power required for delivering information. These features facilitate for \gls{SM} to have a better energy efficiency than that of vectoring.

\section{Simulation results}\label{S5}

\begin{table}[tp!]
\vspace*{1mm}
\caption{Default parameters used for the investigated \gls{DSL} communications.}
\vspace*{-3mm}
\begin{center}
{\rowcolors{1}{green!80!yellow!50}{green!70!yellow!40}
\begin{tabular}{l|l}
\hline\hline
 Occupying frequency                         & $2$ MHz $\sim$ $104.4$ MHz \\ \hline
 Number of tones      $K$                       & 2048 \\ \hline
Bandwidth of a single tone $\Delta_{f}$                    & 0.05 MHz \\ \hline
PSD of additive white Gaussian noise                       & -140 dBm/Hz \\ \hline
Number of groups $N$                               & 2 \\ \hline
Number of lines for each group $M$        & 2  \\ \hline
%Selected frequencies $f_{c}$        & $26.975, 51.975, 76.975, 101.975$ MHz \\ \hline
 Loop length       & 100 m (200 m)\\ \hline
 Modulation order of vectored DSL $J_{\text{vec}}$   & 4 \\ \hline
Supply voltage $V_{s}$   & 4 V\\ \hline
Quiescent current $I_{Q}$   & 11.1 mA \\ \hline
Transformed resistance of the line $R_{\text{line}}^{'}$   & 64 $\Omega$ \\ \hline
Power consumed by the hybrid circuit $P_{\text{Hybrid}}$   & 50 mW \\ \hline
\end{tabular}
}
\end{center}
\label{Table:TAB1}  % Table 1
\vspace*{1mm}
\end{table}

In this section, we investigate the {\color{black}achievable performance of the, CCMC capacity, DCMC capacity, } energy efficiency and the \gls{BER}. The vectoring scheme \cite{cendrillon2006near} is included as benchmark. The direct and crosstalk channels were characterized by the measurements of the UK BT Group plc. The measured channels consist of 2048 subchannels spanning from $2$ MHz to $104.4$ MHz, where each subchannel occupies $0.05$ MHz. Again, the \gls{PSD} of the additive white Gaussian noise (AWGN) is set to $-140$ dBm/Hz {\color{black}by considering all potential noise contributions \cite{acatauassu2012measurement}. The range of power investigated is $3 \text{dBm} \le P_{t} \le 30 \text{dBm}$} allocated for each user over the $2048$ subcarriers. {\color{black} Explicitly, the transmit power $P_{t}$ will be equally shared by the $M$ activated copper pairs of each user, saying $P_{t}^{\text{SM}} = P_{t}$ for the proposed group-based SM transmission and $P_{t}^{\text{vec}} = P_{t} / M$ for the vectoring scheme. The energy efficient is calculated by considering the total power consumption for generating a transmit power that consumed by the \gls{LD}. Again, the \gls{LD}'s power consumption is obtained via Eq.~(\ref{EQ60:LD_power}). The values of the supply voltage, the quiescent current and the transformed resistance of the line are \cite{wolkerstorfer2012modeling}} $V_{s} = 4$ V, $I_{Q} = 11.1$ mA and $R_{\text{line}}^{'} = 64$ $\Omega$, respectively. Furthermore, the power consumed by the hybrid circuit is $P_{\text{Hybrid}} = 50$ mW, as summarized in Table~I. The number of twisted pairs used for each group is $M = 2$. Unless otherwise specified, these default parameter values of Table~I were used throughout the simulations.

{\color{black}
\subsection{Energy efficiency analysis based on CCMC}\label{S5a}
In this subsection, we investigate the achievable CCMC capacity and energy efficiency. The energy efficiency {\color{black} is calculated based on the achievable CCMC capacity} when operating exactly at loop lengths of $100$ m and $200$ m, as shown in Fig.~\ref{FIG5:EE_CCMC_100m} and Fig.~\ref{FIG6:EE_CCMC_200m}, respectively. Explicitly, both the CCMC capacity and the energy efficient share the same x-axis, namely the transmit power\footnotemark\footnotetext{Again, the energy efficient is calculated by considering the total power consumed by the LD that generates the transmit power $P_{t}$.}. The CCMC capacities are represented by the black lines and are quantified on the left-y-axis, while the energy efficient are exhibited by the blue lines measured on the right-y-axis.

It can be seen from Fig.~\ref{FIG5:EE_CCMC_100m} and Fig.~\ref{FIG6:EE_CCMC_200m} that the vectoring scheme outperforms the proposed group-based SM in term of CCMC capacity at the loop length of 100 m and 200m. However, the proposed group-based SM outperforms the vectoring scheme in term of energy efficient when the transmit powers are lower than 15 dBm at the loop length of 100 m and lower than 19 dBm at the loop length of 200m, respectively. Moreover, upon comparing Fig.~\ref{FIG5:EE_CCMC_100m} and Fig.~\ref{FIG6:EE_CCMC_200m}, we can infer that the energy efficiency of all the {\color{black}two schemes} considered becomes worse upon increasing the loop length. Furthermore the proposed group-based \gls{SM} remains the better than the vectoring scheme in a wider range of transmit power.

\begin{figure}[tbp!]
\begin{center}
 \includegraphics[width=1.0\columnwidth,angle=0]{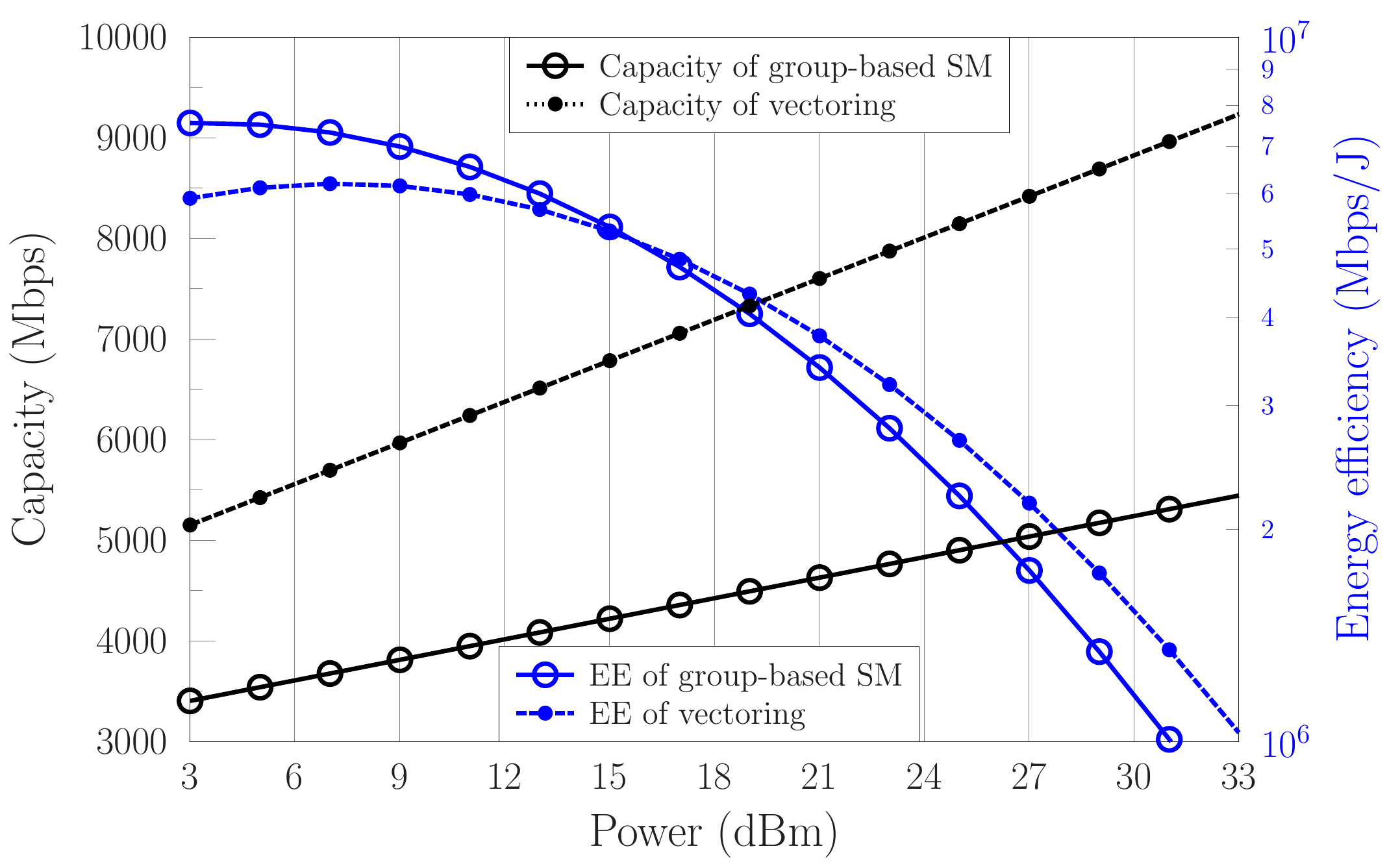}
\end{center}
\vspace{-6mm}
\caption{Energy efficiency when operating exactly at the CCMC capacity at a DSL length of 100 m. The results were calculated by substituting Eq.~(40) and Eq.~(41) into Eq.~(55) and Eq.~(56), respectively.}
\label{FIG5:EE_CCMC_100m}
\end{figure}

\begin{figure}[tbp!]
\begin{center}
 \includegraphics[width=1.0\columnwidth,angle=0]{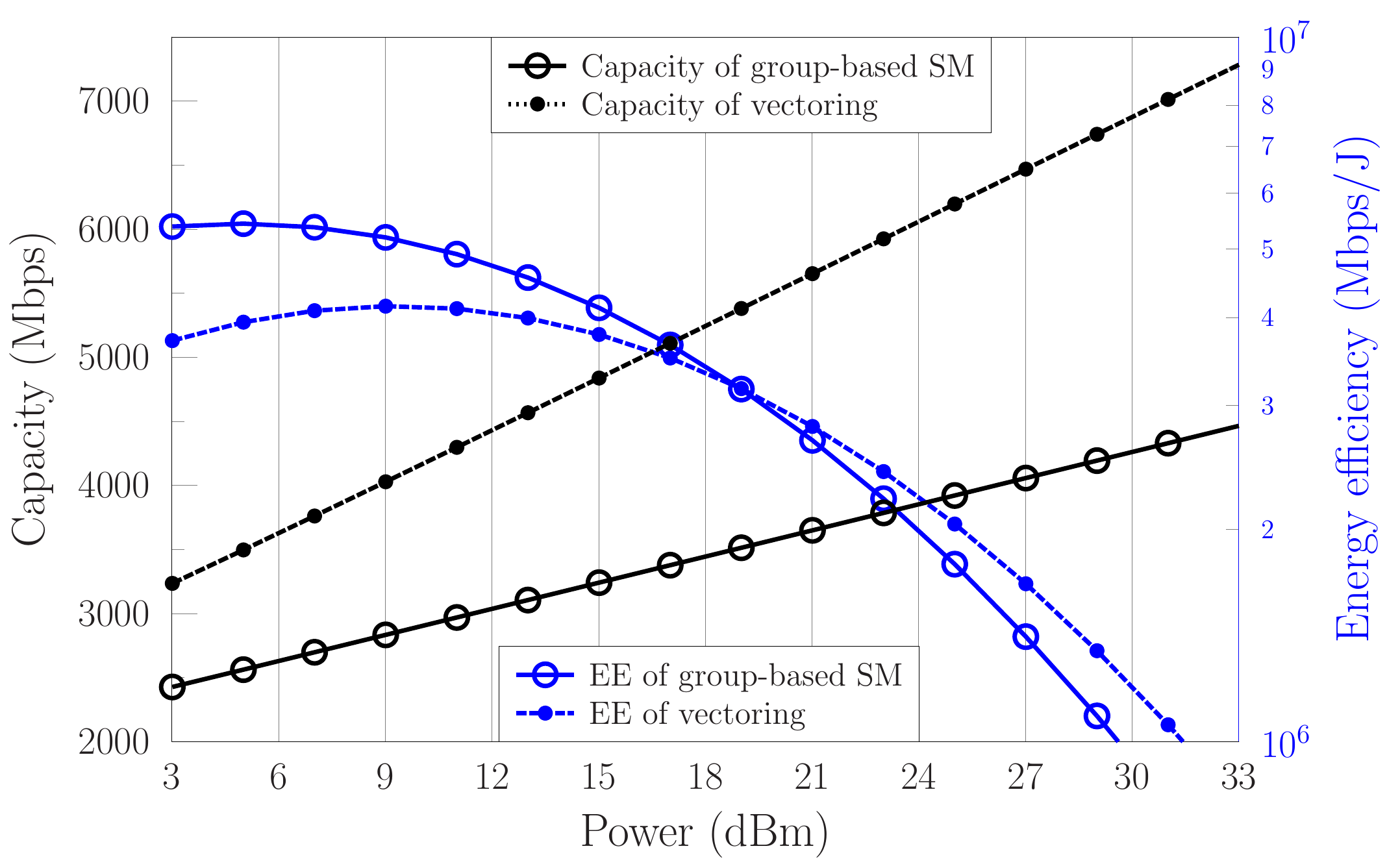}
\end{center} 
\vspace{-6mm}
\caption{Energy efficiency when operating exactly at the CCMC capacity at a DSL length of 200 m. The results were calculated by substituting Eq.~(40) and Eq.~(41) into Eq.~(55) and Eq.~(56), respectively.}
\label{FIG6:EE_CCMC_200m}
\end{figure}

\subsection{Energy efficiency analysis based on DCMC}

The achievable energy efficiency associated with operating at the \gls{DCMC} capacity is investigated in this subsection. 
Firstly, we compare the total capacity of the proposed group-based SM and of the vectoring scheme.
Furthermore, the impact of the loop length, of the number of {\color{black}simultaneous serviced and the number of} the modulation order are investigated in Fig.~\ref{FIG7:EE_DCMC_100m} and Fig.~\ref{FIG8:EE_DCMC_200m}, respectively. {\color{black}The default modulation of the vectoring scheme is 4-\gls{QAM} for the investigation in Fig.~\ref{FIG7:EE_DCMC_100m} and Fig.~\ref{FIG8:EE_DCMC_200m}.} In order to achieve the same bit rate as the vectoring scheme, the default modulation mode of the group-based \gls{SM} is 8-\gls{QAM}. Unless otherwise specified, these default modulations are used throughout this subsection.

\begin{figure}[tbp!]
\begin{center}
 \includegraphics[width=1.0\columnwidth,angle=0]{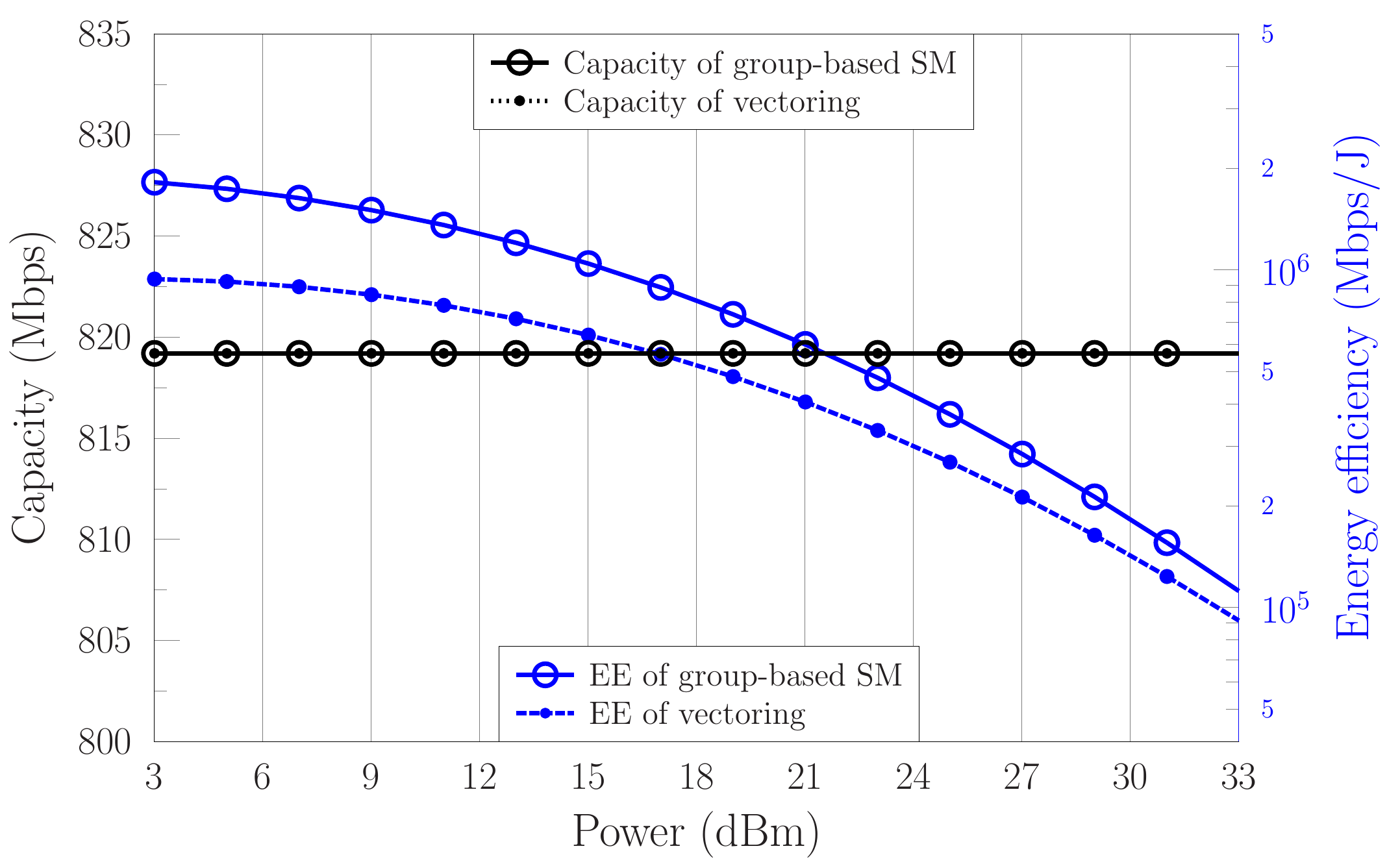}
\end{center}
\vspace{-6mm}
\caption{Energy efficiency when operating exactly at the DCMC capacity at loop length of 100 m. The results were calculated by substituting Eq.~(51) and Eq.~(49) into Eq.~(55) and Eq.~(56), respectively.}
\label{FIG7:EE_DCMC_100m}
\end{figure}

Fig.~\ref{FIG7:EE_DCMC_100m} portrays the energy efficiency when operating exactly at  the \gls{DCMC} capacity at the loop length of 100 m. We can see that the group-based \gls{SM} outperforms the vectoring scheme at all of the examined transmit power range. More specifically, the proposed group-based outperforms the vectoring scheme at about $5.7 \times 10^5$ Mbps/J in term of energy efficient. By contrast, the achievable DCMC capacities of the proposed group-based SM and the vectoring scheme are almost same, both of them have arrived at their saturated capacities of $819.2$ Mbps at the loop length of 100m for the whole examined transmit power range. Furthermore, the energy efficient gap between the proposed group-based SM and the vectoring scheme becomes narrower and narrower upon increasing the transmit power. 

Furthermore, by observing the energy efficiency in Fig.~\ref{FIG8:EE_DCMC_200m}, we can see that the proposed group-based SM remains better than the vectoring scheme in term of energy efficient at the loop length of 200 m. By contrast, the vectoring scheme outperforms the proposed SM in term of DCMC capacity in the transmit power range of $[3 \text{dBm}, 11 \text{dBm}]$m, and both of them achieve almost the same DCMC capacity when the transmit power is higher than 11 dBm.
}

\begin{figure}[tbp!]
\begin{center}
 \includegraphics[width=1.0\columnwidth,angle=0]{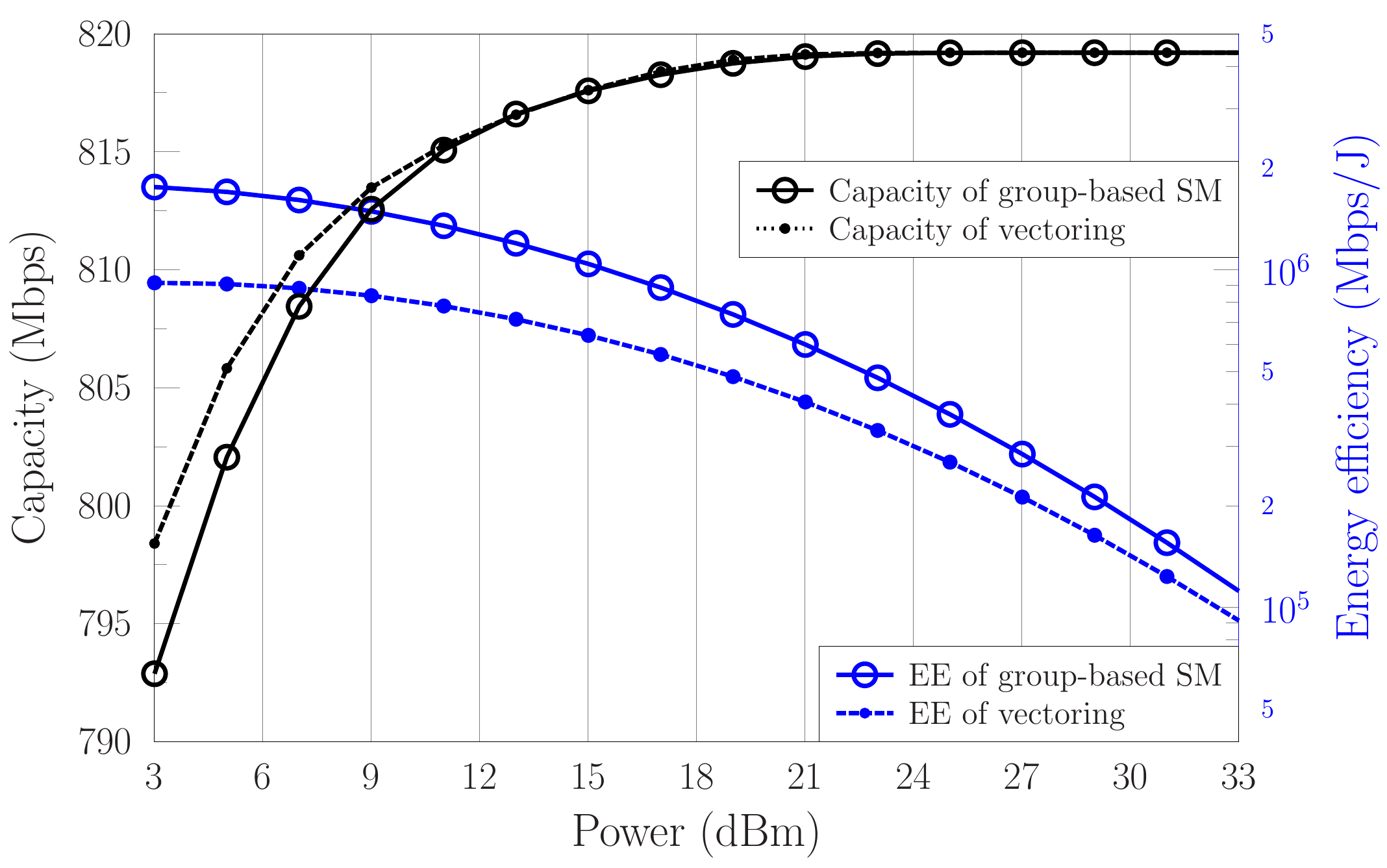}
\end{center}
\vspace{-6mm}
\caption{Energy efficiency when operating exactly at the DCMC capacity at loop length of 200 m. The results were calculated by substituting Eq.~(51) and Eq.~(49) into Eq.~(55) and Eq.~(56), respectively.}
\label{FIG8:EE_DCMC_200m}
\end{figure}

{\color{black}
\subsection{Energy efficiency on a specific tone}
As observed in Fig.~\ref{FIG2:FEXT_channel}, both the direct path and the FEXT have different features at different frequencies (tones). In this subsection, we will investigate the achievable energy efficient on selected Representative tones. The \gls{PSD} of the additive white Gaussian noise (AWGN) is set to $-140$ dBm/Hz. The range of power allocated to each user is in the range of $[0, 20]$ dBm, thus the power allocated to each tone is in the range of $[-33.11, -13.11]$ dBm considering the equally power allocation for all the tones. Again, the energy efficient is calculated by considering the total power consumption for generating a transmit power that consumed by the \gls{LD}, as seen in Eq.~(\ref{EE_k_vec}) and Eq.~(\ref{EE_k_vec}). The selected tone for our investigation are $500$-th, $1000$-th, $1500$-th and $2000$-th tones associated with central frequencies of $f_{c} = 26.975$ MHz, $f_{c} = 51.975$ MHz, $f_{c} = 76.975$ MHz and $f_{c} = 101.975$ MHz. Moreover, we also consider longer loop length of the serviced DSL lines, saying the loop length of 400m. 
}

The energy efficiency achieved when operating exactly at the \gls{CCMC} for a  loop length of $400$ m is shown in Fig.~\ref{FIG9:EE_CCMC_400m}. Observe that the proposed group-based \gls{SM} outperforms the vectoring scheme at all the four subchannels investigated.

\begin{figure}[tbp!]
\begin{center}
 \includegraphics[width=1.0\columnwidth,angle=0]{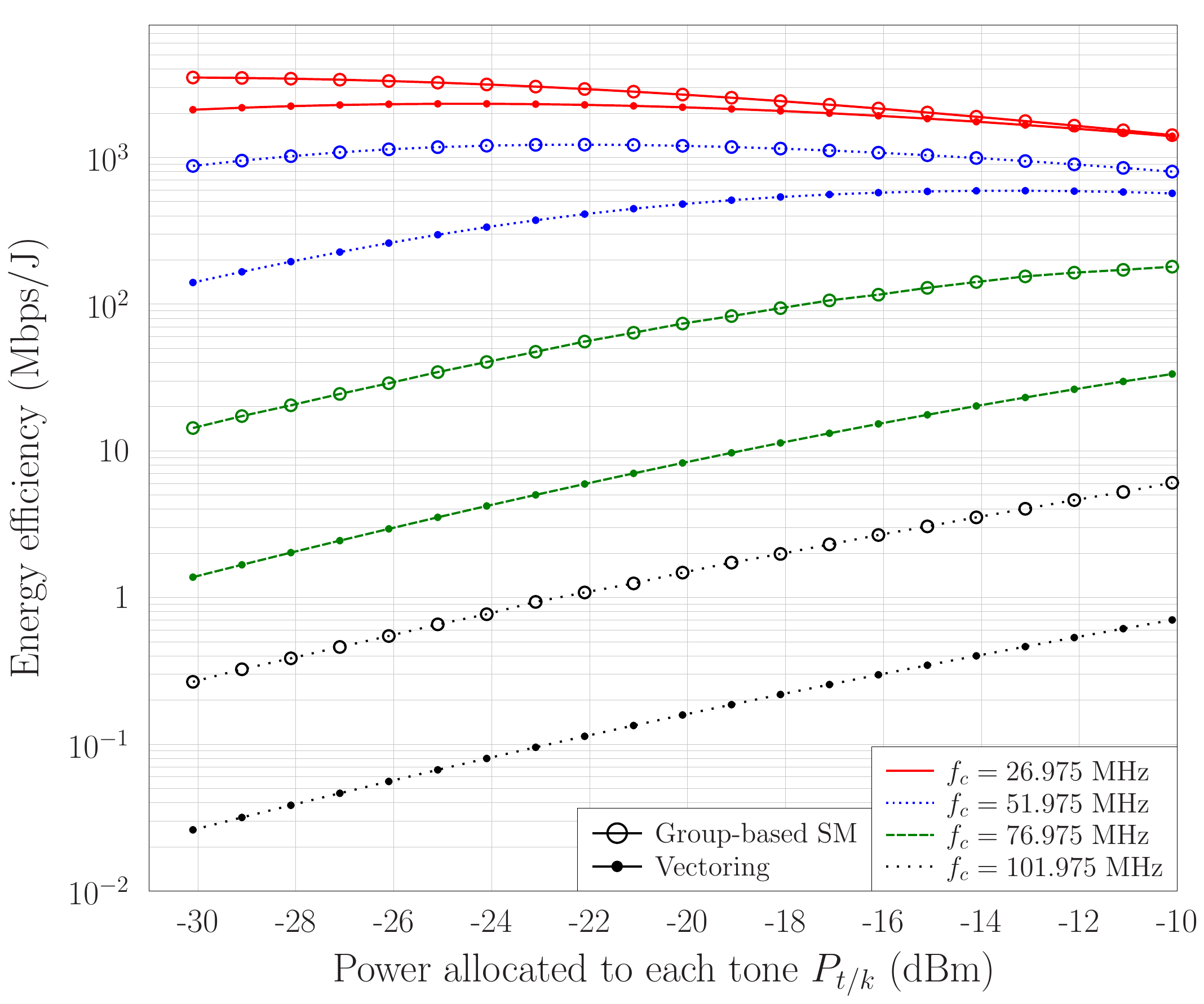}
\end{center}
\vspace{-6mm}
\caption{Energy efficiency when operating exactly at the CCMC capacity at a DSL length of 400 m. The results were calculated by substituting Eq.~(40) and Eq.~(41) into Eq.~(53) and Eq.~(54), respectively.}
\label{FIG9:EE_CCMC_400m}
\end{figure}

\begin{figure}[tbp!]
\begin{center}
 \includegraphics[width=1.0\columnwidth,angle=0]{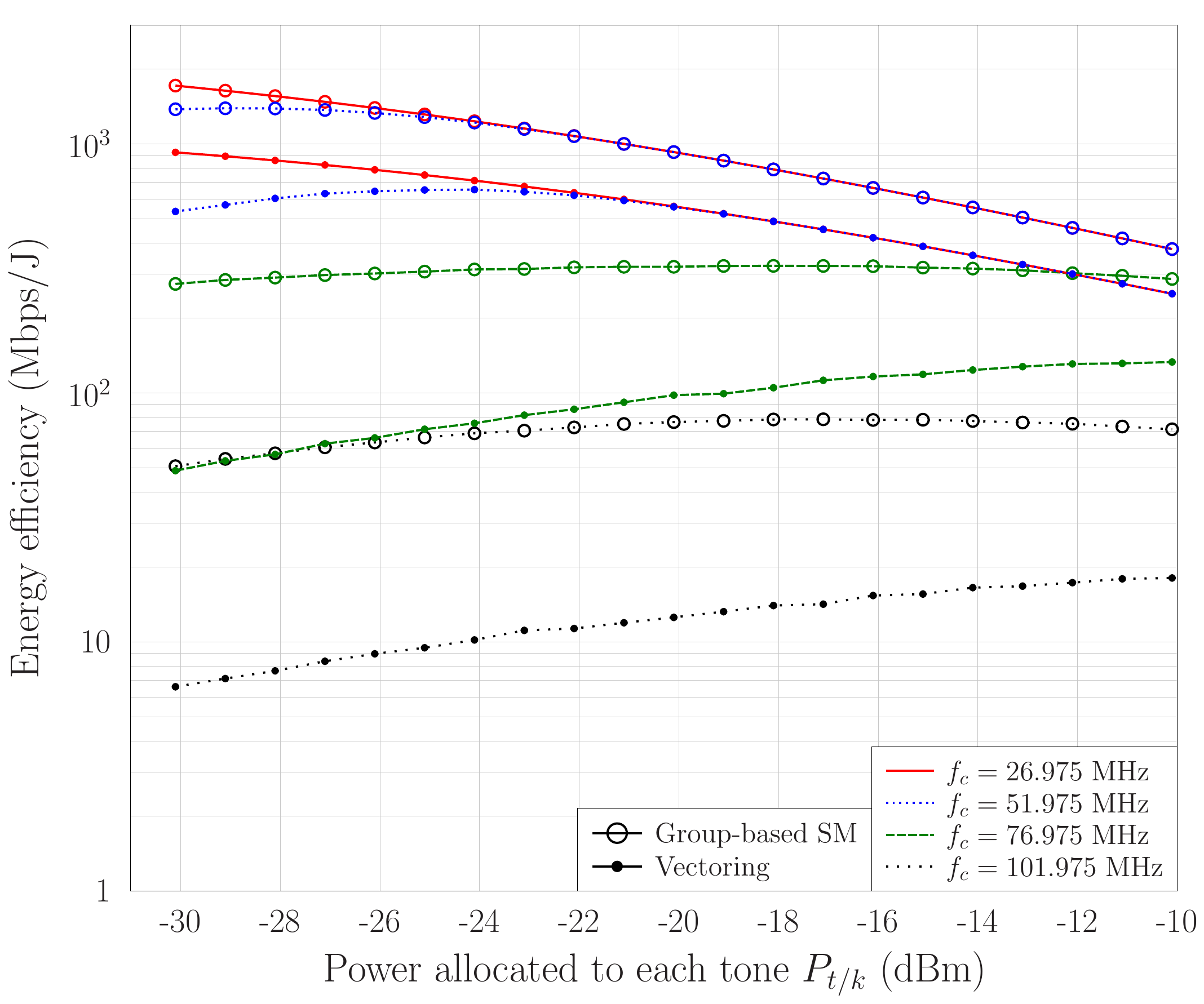}
\end{center}
\vspace{-6mm}
\caption{Energy efficiency when operating exactly at the DCMC capacity at loop length of 400 m. The results were calculated by substituting Eq.~(51) and Eq.~(49) into Eq.~(53) and Eq.~(54), respectively.}
\label{FIG10:EE_DCMC_400m}
\end{figure}

Fig.~\ref{FIG10:EE_DCMC_400m} portrays the energy efficiency when operating exactly at  the \gls{DCMC} capacity at the loop length of 400 m. We can see that the group-based \gls{SM} outperforms the vectoring scheme at the four frequencies of $f_{c} = 26.975$ MHz, $f_{c} = 51.975$ MHz, $f_{c} = 76.975$ MHz and $f_{c} = 101.975$ MHz. More specifically, the normalized energy efficiency of the proposed group-based \gls{SM} at the frequency of $f_{c} = 26.975$ MHz is similar to that at $f_{c} = 51.975$ MHz. The same phenomenon is also observed for the vectoring scheme at frequencies of $f_{c} = 26.975$ MHz and $f_{c} = 51.975$ MHz. Explicitly, the energy efficiency of the proposed group-based \gls{SM} is about 2.5 times higher than that of the vectoring scheme at the frequency of $f_{c} = 26.975$ and loop length of $400$m. By contrast, the energy efficiency of the proposed group-based \gls{SM} is about 1.4 times higher than that of the vectoring scheme at the lower transmission power assigned to the subchannel of $f_{c} = 26.975$ at loop length $400$m. The energy efficiency of the proposed group-based \gls{SM} becomes about 2.5 times higher than that of the vectoring scheme at higher transmission powers.

Intuitively, the crosstalk becomes more serious upon increasing the number of twisted pairs. Hence the impact of crosstalk is investigated by increasing the number of twisted pairs in Fig.~\ref{FIG11:EE_DCMC_400m_N2N3}. Explicitly, in Fig.~\ref{FIG11:EE_DCMC_400m_N2N3}, we compare the achievable energy efficiency of $N = 2$ and $N = 3$ groups, where each group has $M = 2$ twisted pairs. It can be seen from Fig.~\ref{FIG11:EE_DCMC_400m_N2N3} that the proposed group-based \gls{SM} aided transmission has almost the same energy efficiency at frequencies of $f_{c} = 26.975$ MHz and $f_{c} = 51.975$ MHz. However, the energy efficiency of the proposed group-based \gls{SM} aided transmission becomes worse upon increasing the number of groups at frequencies $f_{c} = 76.975$ MHz and $f_{c} = 101.975$ MHz. The vectoring scheme exhibits a better energy efficiency at $N = 2$  groups compared to $N = 3$ groups at all the four frequencies of $f_{c} = 26.975$ MHz, $f_{c} = 51.975$ MHz, $f_{c} = 76.975$ MHz and $f_{c} = 101.975$ MHz. The energy efficiency comparison of Fig.~\ref{FIG11:EE_DCMC_400m_N2N3} between $N = 2$ groups and $N = 3$ groups indicates that the vectoring scheme is more sensitive to the crosstalk.

\begin{figure}[tbp!]
\begin{center}
 \includegraphics[width=1.0\columnwidth,angle=0]{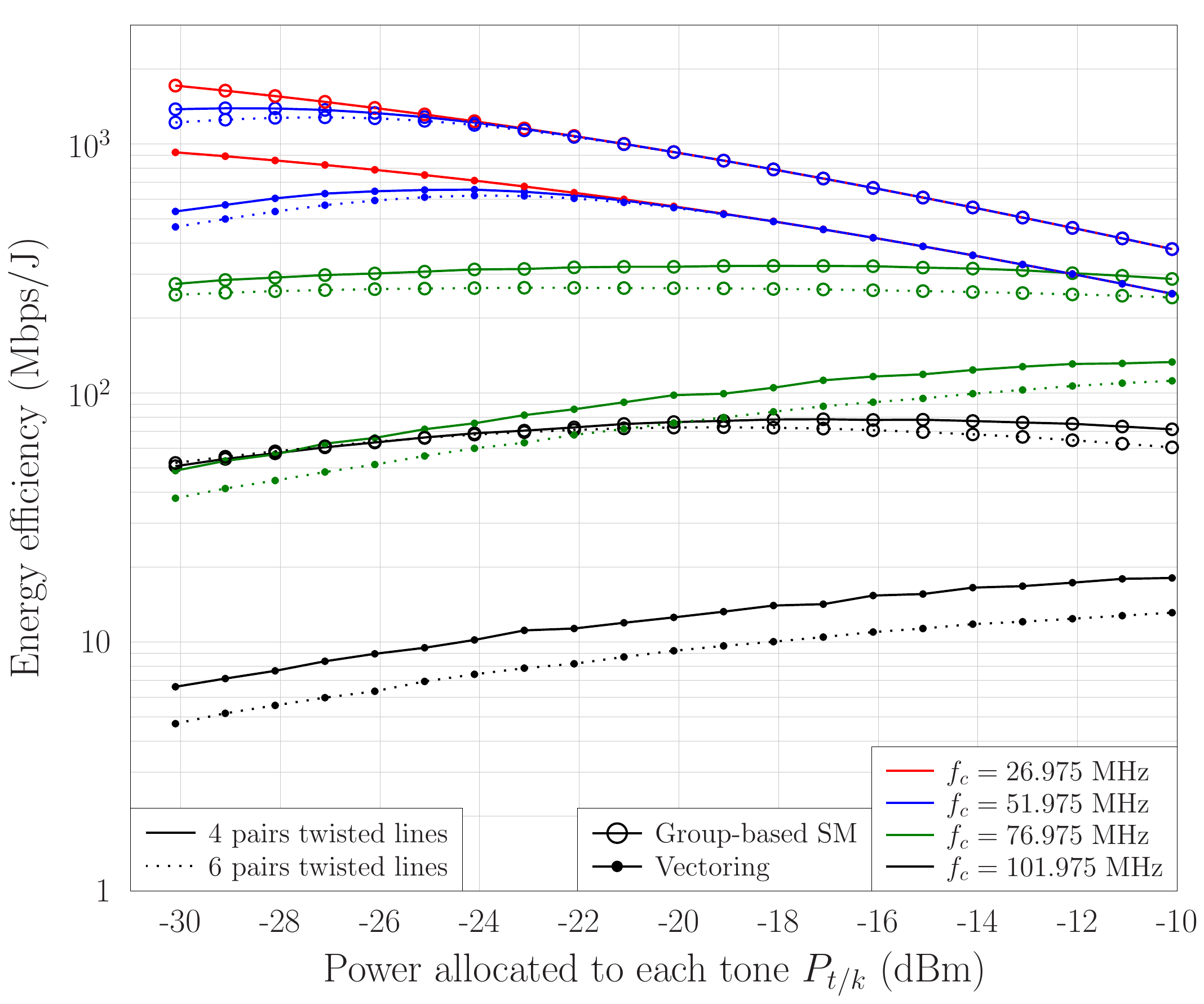}
\end{center}
\vspace{-6mm}
\caption{Comparison of energy efficiency of $(N = 2, M = 2)$ and $(N = 3, M = 2)$ when operating at the DCMC capacity and  at a loop length of 400 m. The results were calculated by substituting Eq.~(51) and Eq.~(49) into Eq.~(53) and Eq.~(54), respectively.}
\label{FIG11:EE_DCMC_400m_N2N3}
\end{figure}

Fig.~\ref{FIG12:EE_DCMC_400m_J4J8} shows the energy efficiency at $\eta = 6$ bits per group use (bpgu), which corresponds to  modulation orders of $J_{\text{vec}} = 8$ and $J_{\text{sm}} = 32$ for vectoring scheme and group-based \gls{SM}, respectively. The energy efficiency at a throughput of $\eta = 4$ bpgu is included as benchmark for investigating the impact of the throughput. It can be seen from Fig.~\ref{FIG12:EE_DCMC_400m_J4J8} that the proposed group-based \gls{SM} shows the best energy efficiency at all the four representative investigated frequencies. The proposed group-based \gls{SM} at $\eta = 6$ bpgu shows a higher energy efficiency than at $\eta = 4$ bpgu. However, the vectoring scheme at $6$ bpgu has a higher energy efficiency than at $\eta = 4$ at frequencies of $f_{c} = 26.975$ MHz and $f_{c} = 51.975$ MHz. By contrast, the vectoring scheme at $\eta = 4$ bpgu has a higher energy efficiency than at $\eta = 6$ bpgu at frequencies of $f_{c} = 76.975$ MHz and $f_{c} = 101.975$ MHz. 

\begin{figure}[tbp!]
\begin{center}
 \includegraphics[width=1.0\columnwidth,angle=0]{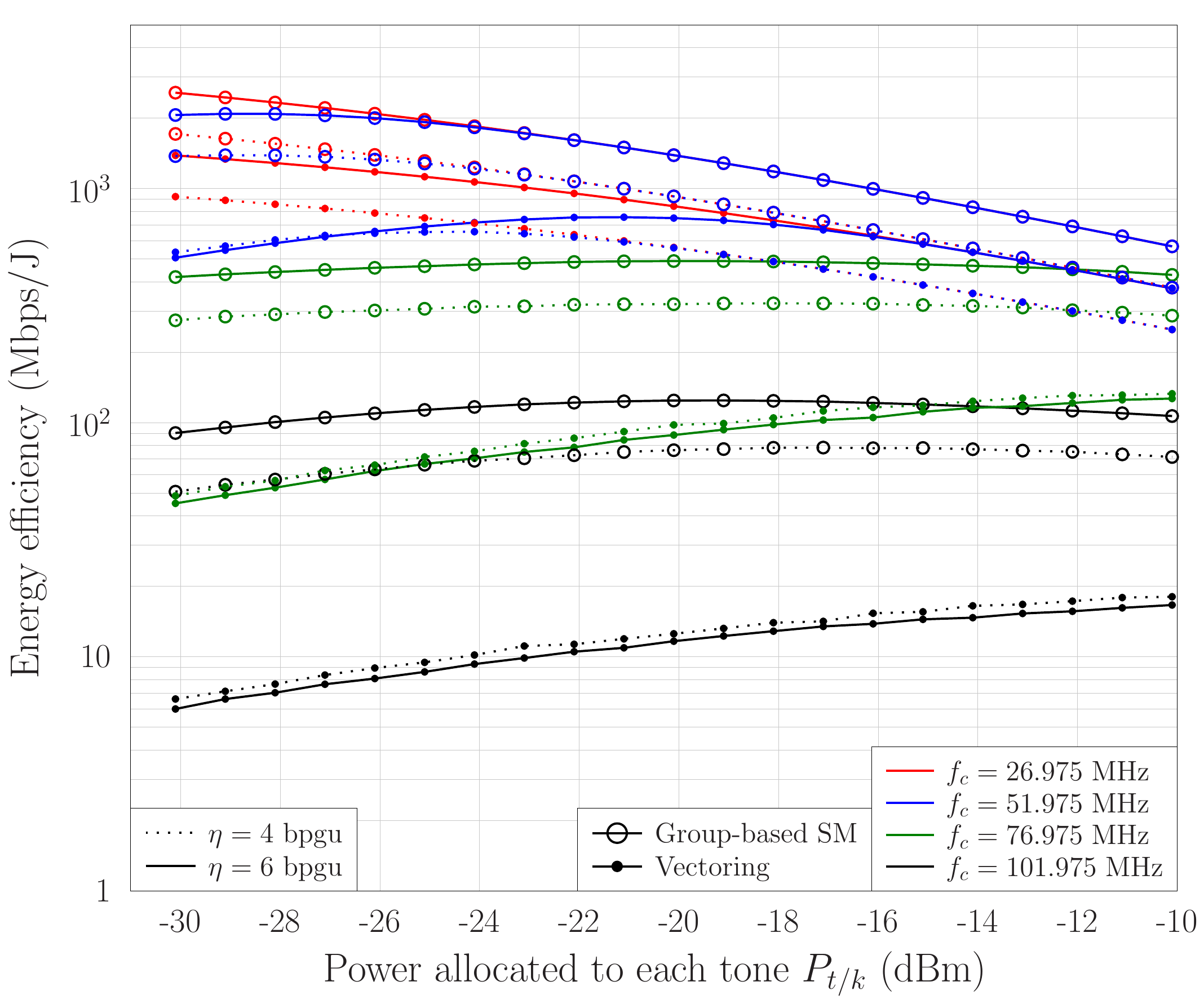}
\end{center}
\vspace{-6mm}
\caption{Comparison of energy efficiency at $\eta = 6$ bpgu based on the DCMC capacity at the loop length of 400m.  The results were calculated by substituting Eq.~(51) and Eq.~(49) into Eq.~(53) and Eq.~(54), respectively.}
\label{FIG12:EE_DCMC_400m_J4J8}
\end{figure}

\subsection{Comparison of BER performance}
The achievable \gls{BER} performance is investigated at the setting of transmit $\text{PSD} = -70$ dBm/Hz for each group, which corresponds to the total transmit power of $P_{t}^{\text{total}} = 10.10$ dBm over $2048$ subchannels for each group, where each subchannel occupies $0.05$ MHz. Note that the total transmit power is shared by $M$ lines for a fair comparison.

Since the \gls{DSL} channel qualities seen in Fig.~\ref{FIG2:FEXT_channel} are quite different at low and high frequencies, we will investigate the achievable \gls{BER} performance at a particular frequency {\color{black}first, and then we investigate the achievable BER performance upon increasing the bandwidth to the full bandwidth occupation of 102.4 MHz.} The selected frequencies are $f_{c} = 26.975$ MHz, $f_{c} = 51.975$ MHz, $f_{c} = 76.975$ MHz and $f_{c} = 101.975$ MHz, which span from low frequency to high frequency.

It can be seen from Fig.~\ref{FIG13:BER_H4_J4} that the proposed SOSD-I outperforms vectoring scheme at the frequencies of $f_{c} = 26.975$ MHz, $f_{c} = 51.975$ MHz and $f_{c} = 76.975$ MHz. Explicitly, at the \gls{BER} level of $1 \times 10^{-6}$, the loop length gains achieved are about $37.5$ m, $25$ m and $9.4$ m at $f_{c} = 26.975$ MHz, $f_{c} = 51.975$ MHz and $f_{c} = 76.975$ MHz, respectively. This may be deemed to be almost identical, but bear in mind that this is achieved at a factor $1/M = 1/2$ lower energy per group. Similarly, the reduced-complexity SOSD-II also outperforms the vectoring scheme at $f_{c} = 26.975$ MHz and $f_{c} = 51.975$. By contrast, at $f_{c} = 76.975$ MHz and $f_{c} = 101.975$ MHz, the proposed SOSD-II advocated only outperforms the vectoring scheme when the loops  are longer than $248$ m and $302$ m, respectively, as reflected by the cross-over of their curves in Fig.~\ref{FIG13:BER_H4_J4}. In order to clearly illustrate the gain achieved gain at a specific BER of $10^{-6}$, we summarize the DSL lengths of the three schemes considered in Table.~\ref{Table:TAB2}.

%\begin{figure*}[tbp!]
\begin{table*}[tp!]
\vspace*{1mm}
\caption{Supported transmission length when the achieved BER is at $10^{-6}$.}
\vspace*{-3mm}
\begin{center}
{
\begin{tabular}{|l|c|c|c|c|}
\hline\hline
 \backslashbox{Schemes}{Subchannels} & $f_{c} = 26.975$ MHz & $f_{c} = 51.975$ MHz & $f_{c} = 76.975$ MHz & $f_{c} = 101.975$ MHz \\
\hline
SOSD-I & 634.4 m  & 423.1 m & 300 m & 206.3 m \\
\hline
SOSD-II & 600.0 m  & 407.2 m & 281 m & 212.5 m \\
\hline
Vectored DSL & 605.0 m  & 400.0 m & 293.0 m & 234.4 m \\
 \hline
\end{tabular}
}
\end{center}
\label{Table:TAB2}  % Table 1
\vspace*{1mm}
\end{table*}
%\end{figure*}[tbp!]

\begin{figure}[tbp!]
\begin{center}
 \includegraphics[width=1.0\columnwidth,angle=0]{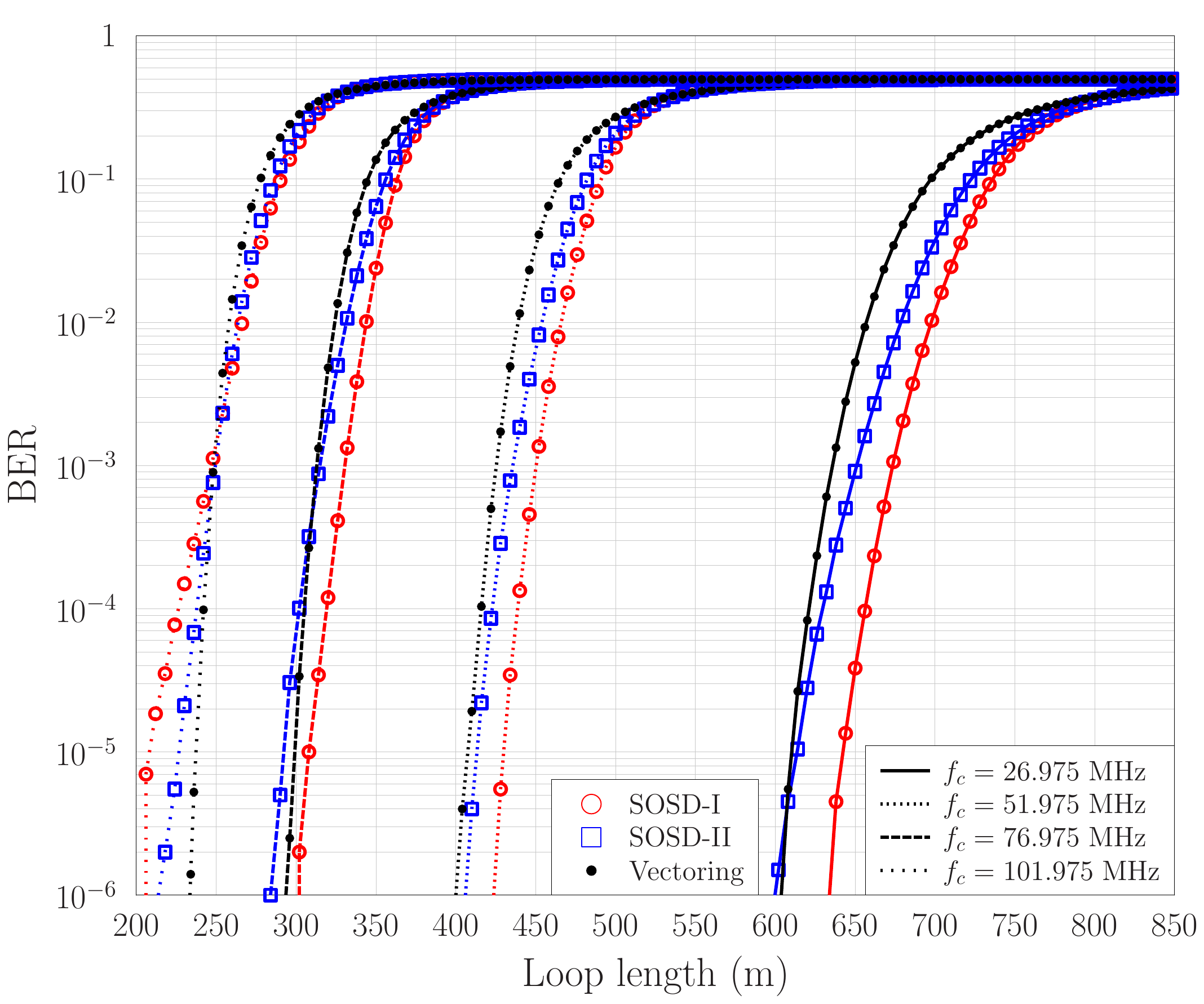}
\end{center}
\vspace{-6mm}
\caption{BER performance versus loop length of group-based SM and vectored DSL. In the upstream multi-user DSL system considered, the number of group is $N = 2$, each group has $M = 2$ twisted pairs. The bit rate is 4 bpgu per group, requiring that $J_{\text{vec}} = 4$ and $J_{\text{sm}} = 8$.}
\label{FIG13:BER_H4_J4}
\end{figure}

\subsection{Overall performance}
In this subsection, we investigate the overall achievable average BER (ABER) attained upon increasing the bandwidth, as shown in Fig.~\ref{FIG14:cumsum_capacity_BER} and Fig.~\ref{FIG15:BER_VS_length}. It can be seen from Fig.~\ref{FIG14:cumsum_capacity_BER} that both the proposed SOSD-I and the reduced
complexity SOSD-II detection schemes exhibits a lower BER than the vectoring scheme right across the entire bandwidth examined at a DSL length of 400m. The same trend is also valid at the DSL length of 600 m.

\begin{figure}[tbp!]
\begin{center}
 \includegraphics[width=1.0\columnwidth,angle=0]{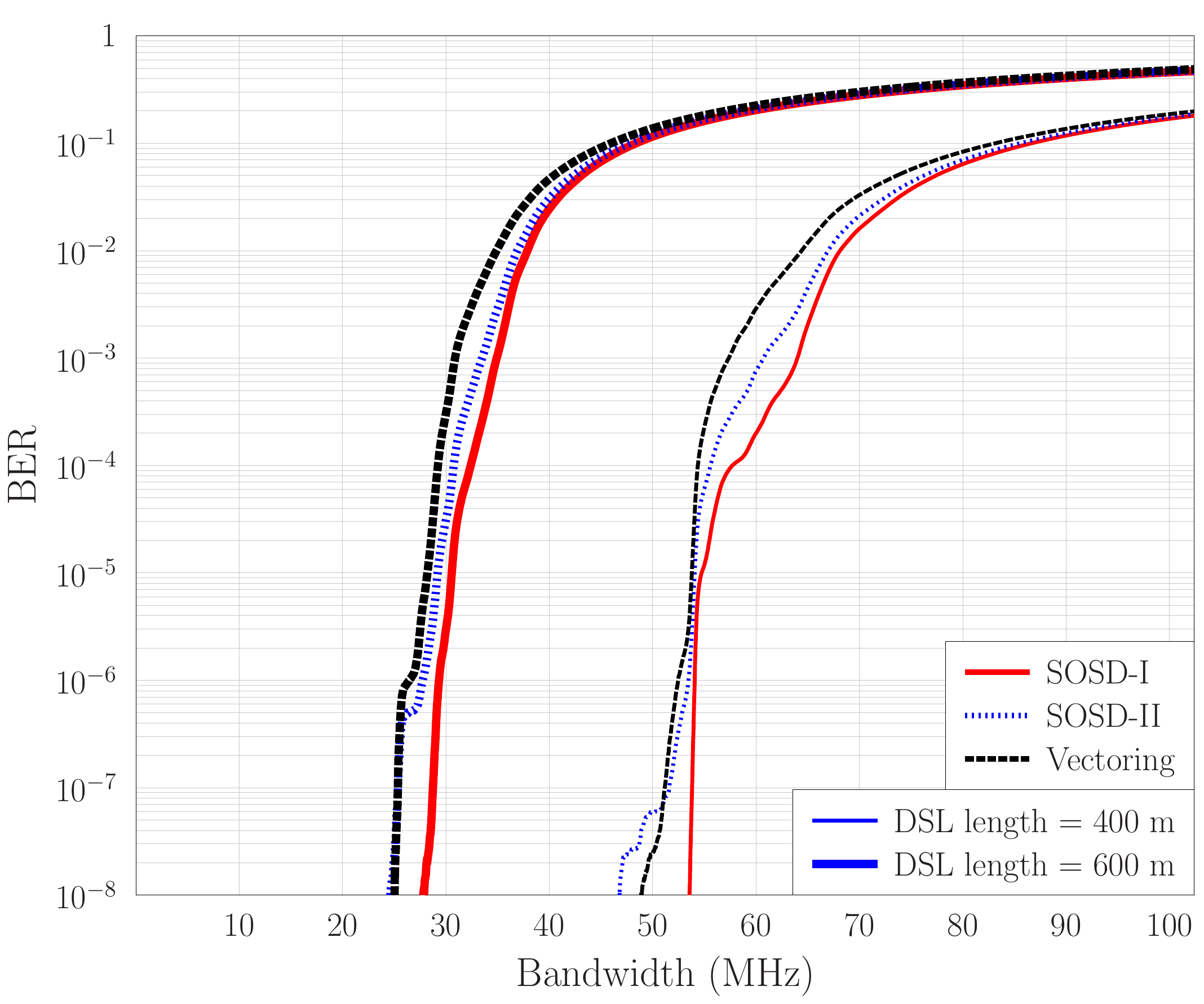}
\end{center}
\vspace{-6mm}
\caption{Comparison of the total achievable capacity and the average BER upon increasing the bandwidth.}
\label{FIG14:cumsum_capacity_BER}
\end{figure}

\begin{figure}[tbp!]
\begin{center}
 \includegraphics[width=1.0\columnwidth,angle=0]{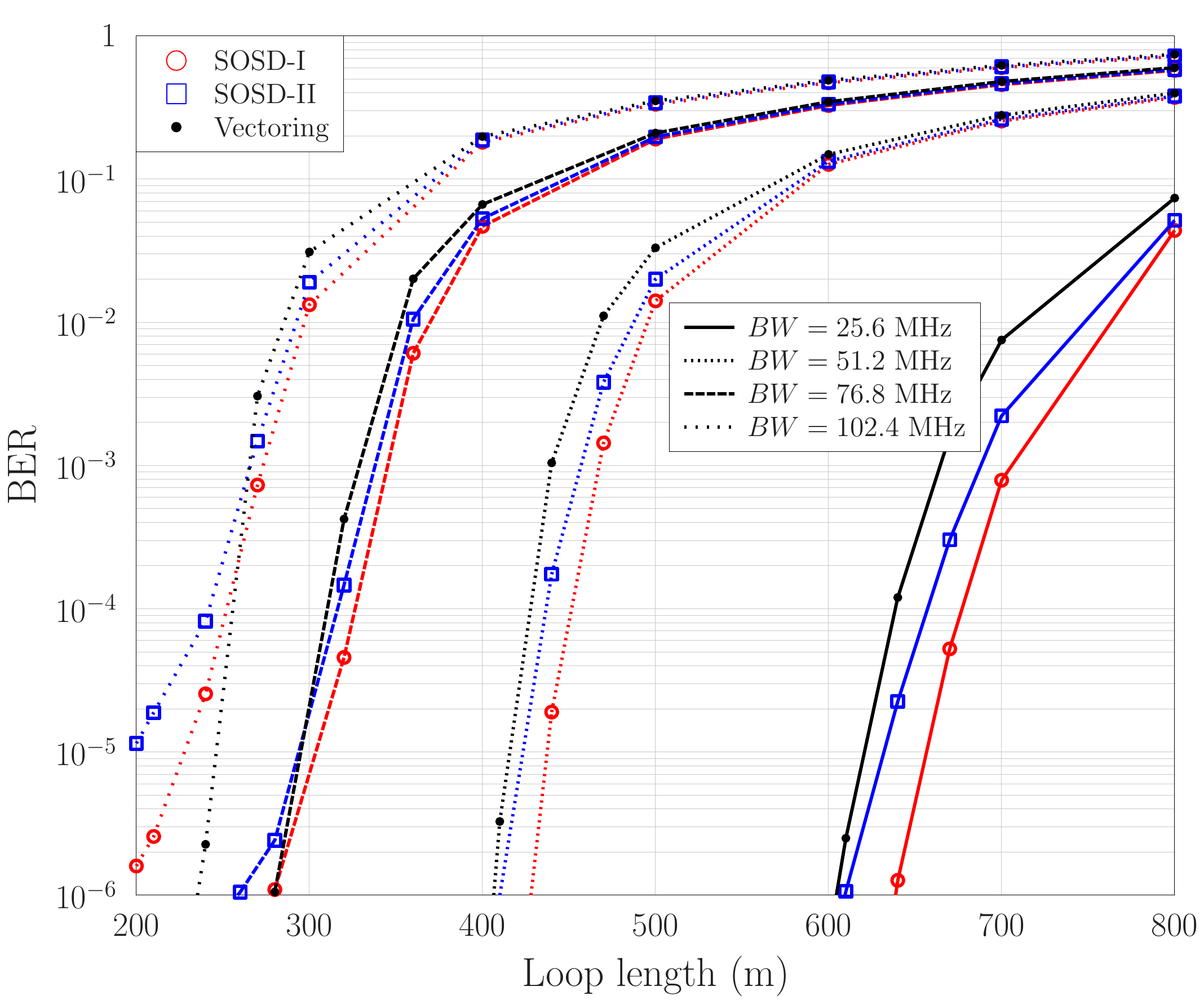}
\end{center}
\vspace{-6mm}
\caption{Comparison of the average BER upon increasing the DSL length.}
\label{FIG15:BER_VS_length}
\end{figure}

The overall BER performance upon increasing the cable length is quantified in Fig.~\ref{FIG15:BER_VS_length}. The proposed SOSD-I is capable of outperforming the vectoring scheme at the bandwidths of $BW = 25.6$ MHz, $BW = 51.2$ MHz and $BW = 76.8$ MHz.  However, at the bandwidth of $BW = 102.4$ MHz, the SOSD-I only outperforms the vectoring scheme, when the DSL length becomes longer than 270 m. By contrast, the reduced complexity SOSD-II detection scheme is capable of outperforming the vectoring scheme at a bandwidth $BW = 25.6$ MHz and bandwidth of $BW = 51.2$ MHz. However, it only outperforms the vectoring scheme, when the DSL lengths are above 300 m and 270 m for the bandwidth of $BW = 76.8$ MHz and $BW = 102.4$ MHz, respectively. The observation of  Fig.~\ref{FIG15:BER_VS_length} shows that the SOSD-I and SOSD-II detection schemes have a better overall BER at longer DSL lengths. By contrast, the vectoring scheme only shows a better BER performance at shorter DSL lengths below 300 m and 270 m at bandwidths of $BW = 25.6$ MHz and of $BW = 51.2$ MHz, respectively.

\section{Conclusions}\label{S6}
In this paper, we proposed a group-based \gls{SM} scheme for the multi-user upstream of \gls{DSL} systems. Explicitly, {\color{black}we consider a emerging bonding scenario that customers have been equipped with two copper pairs.} The twisted pairs are indexed and divided into groups, where each group only activates a single twisted pair in order to reduce the cross-talk and hence indirectly also the power consumption. This is achieved by delivering ``\textit{virtual bits}" via identifying the index of the activated channel of each group. Furthermore, a pair of sub-optimal soft turbo detection schemes were proposed by exploiting the \gls{CWDD} property of the \gls{DSL} channel. The achievable energy efficiency was investigated, when operating exactly at the \gls{CCMC} and \gls{DCMC} capacities. Explicitly, the impact of loop length, the bit rate and of the binder channels was investigated by simulations. The proposed group-based \gls{SM} exhibits the best energy efficiency for all the examined transmit powers and DSL loop lengths. Furthermore, the overall average BER were investigated upon increasing bandwidth, which showed that the group-based SM is capable of outperforming the vectoring scheme {\color{black}at lower frequencies and longer DSL loop lengths.} Moreover, our investigations also showed that the vectoring scheme exhibited a better performance at very short DSL length, while both the proposed SOSD-I and the SOSD-II began to outperform the vectoring scheme for DSL lengths beyond 300 m and 270 m. This observation suggests the feasibility of a hybrid transmission scheme by exploiting the specific DSL channel characteristics in terms of its cable length and frequency as well as bandwidth used.

\bibliographystyle{IEEEtran}

\end{document}